\RequirePackage{fix-cm}
\documentclass{svjour3}                     
\smartqed  
\usepackage{graphicx}
\usepackage{fancybox}
\usepackage{multirow}
\usepackage{xspace}
\usepackage{color}
\usepackage{xcolor}
\usepackage{framed}
\usepackage{threeparttable}
\usepackage{subfigure}
\usepackage{listings}
\usepackage{booktabs}
\usepackage{natbib}
\usepackage{verbatim}
\usepackage{hyperref}
\usepackage{makecell}

\newcommand{\ie}{\textit{i}.\textit{e}.}
\newcommand{\eg}{\textit{e}.\textit{g}.}

\newcommand{\new}[1]{\textcolor{black}{#1}}
\newcommand{\rev}[1]{\textcolor{black}{#1}}
\newcommand{\approach}{Zecoler\xspace}

\journalname{ArXiv}

\begin{document}
\title{
Zero-Shot Code Representation Learning via Prompt Tuning

}

\titlerunning{Zero-Shot Code Representation Learning}        

\author{Nan Cui \and 
        Xiaodong Gu \and
        Beijun Shen
}

\institute{Nan Cui, Xiaodong Gu, Beijun Shen \at
              School of Software, Shanghai Jiao Tong University, China \\
\email{\{cuinan, xiaodong.gu,  bjshen\}@sjtu.edu.cn}           
}

\date{2024/4/12}

\maketitle

\begin{abstract}
Learning code representations has been the core prerequisite of many software engineering tasks such as code clone detection and code generation. \rev{State-of-the-art program representation techniques mainly utilize pre-trained language models (PLMs) such as CodeBERT. A Transformer encoder is firstly pre-trained on a large-scale code corpus to acquire general knowledge about source code. The pre-trained model is then fine-tuned on specific tasks using an amount of labeled data}. However, gathering training samples for the downstream tasks can be prohibitively expensive and impractical for domain-specific languages or project-specific tasks. \new{Besides, pre-training and downstream tasks are usually heterogeneous, which makes it difficult to fully explore the knowledge learned during pre-training.}
In this paper, we propose \approach, a zero-shot approach for learning code representations. \approach is built upon a pre-trained programming language model. In order to elicit knowledge from the PLMs efficiently, \approach casts the downstream tasks to the same form of pre-training objectives by inserting trainable prompts into the original input. These prompts can guide PLMs on how to generate better results.
Subsequently, we employ the prompt tuning technique to search for the optimal prompts for PLMs automatically. This enables the representation model to efficiently fit the downstream tasks through fine-tuning on the dataset in source language domain and then reuse the pre-trained knowledge for the target domain in a zero-shot style.
We evaluate \approach in \new{five code intelligence tasks including code clone detection, code search, method name prediction, code summarization, and code generation.}
We experiment in multiple programming languages without giving labeled samples, \eg, Solidity and Go, with model trained in corpora of common languages such as Java. The results show that our approach significantly outperforms baseline models under the zero-shot setting. \new{For example, the accuracy of code search is improved by 30\% compared to fine-tuning. In addition, qualitative analysis demonstrates its superior generalizability under both cross-lingual and monolingual few-shot settings.}


\end{abstract}

\section{Introduction}
\label{sec:intro}
\new{Deep learning models have been widely applied to a variety of software engineering tasks, such as clone detection~\citep{FangLS0S20}, code summarization~\citep{ChoiBNL21}, and code search~\citep{HaldarWXH20}. In order to apply deep learning to these tasks, source code needs to be represented as vectors that reflect their deep semantics.}
For example, in the clone detection task, code representations can be used to identify similar features between two code snippets~\citep{ZhangHZWLS21}.

\new{Pre-trained language models (PLMs) of code such as CodeBERT~\citep{codebert}, CodeT5~\citep{codet5}, and PL-BART~\citep{plbart} have been the cutting-edge code representation technology. A code PLM is pre-trained to learn code representations on large-scale code corpora with self-supervised objectives, and then is fine-tuned to adapt to downstream tasks.
They have demonstrated a better understanding of the semantics of source code than previous deep learning models such as code2vec~\citep{code2vec} and ASTNN~\citep{zhang2019astnn}.
}

\smallskip\noindent\new{\textbf{Challenges.}} 
Despite showing promising results, \new{fine-tuning a PLM on a specific task is challenging. 
First, its performance relies on the availability of sufficient training data for downstream tasks. However, in practice, the labeled data for downstream tasks is far limited, especially for domain-specific languages or project-specific tasks.} For example, Solidity is a new language that is specifically designed for smart contracts. 
Labeling Solidity code requires much domain knowledge on Blockchain, which is often costly and laborious.
Also, the collected data is significantly redundant~\citep{ChenLZ0Z21}. 
This restricts the collection of supervised data and causes poor representations learned by the model~\citep{codenet}.

\new{Second, the pre-training tasks (e.g., masked language model (MLM)) are usually heterogeneous from downstream tasks such as code search. } 
As such, the reusability of prior knowledge learned in the pre-training phase may be limited in the fine-tuning phase.
This is even more challenging when there is no or insufficient training data for downstream tasks in domain-specific languages. Large PLMs can easily overfit scarce data, which lead to poor task fitting. 
\new{Hence, an efficient mechanism to elicit knowledge from PLMs to the downstream tasks in the zero-shot scenario is highly demanded.}


\smallskip\noindent\new{\textbf{Our work.}} 
In this paper, we propose \approach (\textbf{Ze}ro-shot \textbf{co}de representation \textbf{le}a\textbf{r}ning), a novel approach for learning code representations for a language that has no labelled data samples. 
\new{The key idea is to transfer the representations of a programming language with sufficient data (i.e., source language such as Java) into a target language that has few training samples (e.g., Solidity).}
Specifically, we adapt prompt-based learning, a new learning paradigm for PLMs, that continually trains a PLM in the source language and then transfers the model to tasks in the target language.

\new{First, by accompanying trainable prompt tokens with the PLM input, \approach adapts the downstream task to the same form as that in pre-training. 
For example, code clone detection can be converted to an MLM task by inserting prompt and ``[MASK]" tokens into the input. 
Then, to optimize the prompt, we learn a continuous task-specific vector on the dataset of downstream task in source language.
In this way, the model can be adapted to the target language and the prompts further guide it to efficiently elicit knowledge of programming languages learned during pre-training. }

To evaluate the proposed approach, we experiment on \new{five classification and generative tasks, including code clone detection, code search, method name prediction, code summarization, and code generation}. The results show that our approach is substantially effective in zero-shot learning of code representations. The accuracy of the three classification tasks in Solidity is 79.8\%, 67.1\%, and 68.1\%, respectively, which is around 14.7\% greater than the strong baseline CodeBERT. \new{The two generative tasks also gain visible improvement in terms of BLEU/CodeBLEU and Rouge-L. 
}

\new{This paper extends our preliminary study, which appears as a research paper in ICPC~\citep{ZeroShot}. In particular, we extend our preliminary work in the following directions:
\begin{itemize}
    \item[1.] \new{We apply zero-shot learning of code representations to code generative tasks that can be modeled by the sequence-to-sequence framework.}
    \item[2.] \new{We provide a more in-depth empirical study to investigate the effectiveness of our approach in code summarization and code generation tasks.} 
\end{itemize}
}

The main contributions of this paper, \new{as a super-set of our preliminary study}, are summarized as follows:
\begin{itemize}
    \item To the best of our knowledge, we are the first to propose zero-shot learning of code representations, which does not require \new{task-relevant training data of the target language for fine-tuning}. 
    \item We propose a prompt-based learning method for zero-shot code representation \new{that can generalize to both classification and generative tasks}. 
    \item We conduct extensive experiments to evaluate the proposed approach \new{on five code intelligence tasks}. Results show that our approach significantly outperforms the baseline models.
\end{itemize}

\new{The rest of the paper is organized as follows. Section 2 provides background knowledge on pre-trained models and zero-shot learning. Section 3 presents the technical details of \approach. The experimental setup is explained in Section 4 and the results are analyzed in Section 5. We discuss the threats to the validity in Section 6 and compare \approach against related work in Section 7. Finally, we conclude our study and mention future work in Section 8.}

\section{Background}
\label{sec:background}
\subsection{Pre-trained Language Models for Code}
\label{sec:pretrain}

Pre-trained language models (PLMs) such as BERT~\citep{DevlinCLT19}, GPT, and T5~\citep{RaffelSRLNMZLL20} \new{have been shown to provide large improvements for a range of natural language tasks. 
The key idea of PLMs is to train a large model on vast corpora and use the resulting representations on tasks for which only limited amounts of labeled data is available.}

A PLM is pre-trained on a large-scale text corpora through a series of self-supervised learning tasks, \eg, masked language modeling (MLM) and next sentence prediction (NSP). 
The MLM task masks a random portion of tokens in the input text and tries to predict the masked words, while the NSP task predicts whether or not two given input segments are coherent. 
\new{These pre-training tasks enable PLMs to learn the general knowledge from large language corpora.} 
Then the PLM is fine-tuned on a task-specific dataset. A fine-tuning header on top of the PLM is optimized via supervised learning tasks in a specific domain. \new{This header is usually a trainable neural network such as a \rev{multi-layer perceptron} (MLP) for classification tasks or a Transformer decoder for generative tasks.}

Given the great success of PLMs in NLP, researchers also seek the adaptations of PLMs for programming languages~\citep{codebert,codet5,codex}.
They customize the pre-training objectives using programming related tasks on a big code corpus.
PLMs have been successfully used to learn code representations and further be adapted in a variety of code intelligence tasks, \new{\eg, clone detection~\citep{FangLS0S20}, code summarization~\citep{ChoiBNL21}, and code search~\citep{HaldarWXH20}}.

\begin{figure}[!t]
    \centering
    \includegraphics[width=0.85\textwidth, clip]{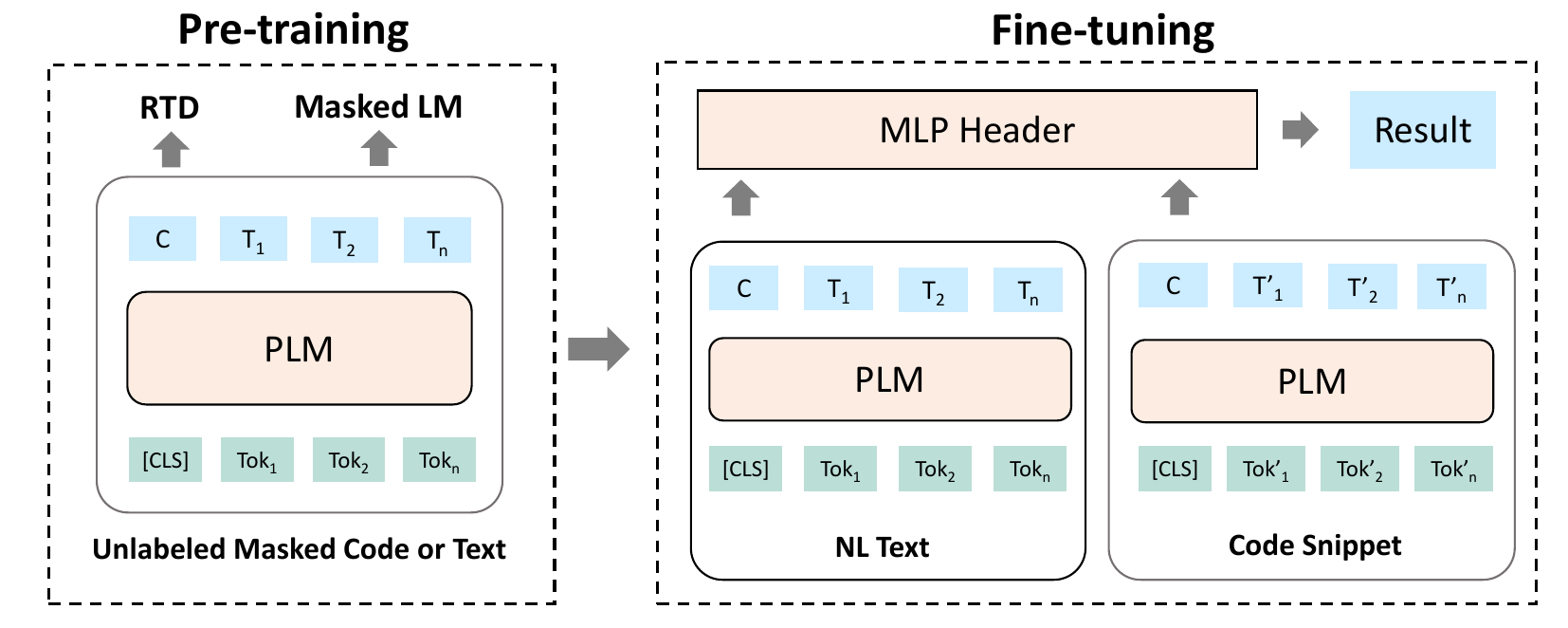}
    \caption{Illustration of the pre-trained code model.}
    \label{fig:plm}
\end{figure}

For example, CodeBERT is built on RoBERTa~\citep{roberta} and is pre-trained with both natural and programming languages. Figure~\ref{fig:plm} illustrates the main pipeline of CodeBERT. The model is first pre-trained on two tasks, namely, MLM and replaced token detection (RTD). The MLM task randomly masks the tokens in natural language and programming language (NL-PL) pairs and trains the model to predict the original words. The RTD task trains the model to detect whether a given token is original or generated by the model. 
Having pre-trained, the pre-trained model is fine-tuned on data of downstream tasks such as clone detection and code search. A fine-tuning header is added to the PLM and is optimized with downstream tasks. 

\subsection{Zero-Shot Learning}
The standard supervised learning approaches train a model with large-scale labelled samples. However, in many tasks such as recognizing name of a new brand or translating a new language, obtaining sufficient training samples is laborious and often impracticable.
Zero-shot learning transfers a learned model \new{from a source domain} to a target domain that has no labelled data, and hence alleviates this ``data hungry'' problem. 
It can be realized through a variety of techniques such as data augmentation~\citep{BorneaPRFS21}, meta-learning~\citep{metalearning1, metalearning2}, PLMs~\citep{DevlinCLT19}, and prompt-based learning~\citep{gpt3}.

\textit{Data augmentation.}
A direct technique of zero-shot learning is data augmentation, namely, enlarging the data set (e.g., randomly inserting samples and noise) so that the model can have sufficient data samples for training~\citep{BorneaPRFS21}. 

\textit{Meta-learning.} 
Another popular strategy for zero-shot learning is meta-learning. Meta learning is also known as ``learning to learn'', which aims at training a meta learner which learns the update rules of the target model~\citep{GuWCLC18}. This enables a machine learning model to achieve competitive performance even with scarce data. However, meta-learning focuses on learning strategies instead of representations. Hence it will be difficult to be generalized across different code intelligence tasks. 


\textit{Pre-trained Language Models.} 
PLMs are pre-trained on large-scale text corpora to learn common knowledge of the languages, and can be generalized to specific tasks with only a few training examples. 
However, PLMs need a fine-tuning phase which continually train the pre-trained model on the downstream tasks. Fine-tuning requires the availability of manually labelled datasets which is laborious and expensive.

\textit{Prompt-based Learning.} 
 To alleviate the data hungry problem of fine-tuning, GPT3~\citep{gpt3} introduces prompt-based learning, a lightweight alternative of fine-tuning for PLMs. 
 \new{A prompt is usually a piece of text inserted in the input to guide the pre-trained model to generate desirable results. For example, one can prepend ``TL;DR'' followed by a few examples in the input of GPT3 to let it summarize the input.}
 
 Unlike fine-tuning, which adds fine-tuning header and re-optimizes the PLM using downstream tasks, the prompt-based learning approach converts downstream tasks (e.g., method name detection) to the same form as the pre-training tasks (e.g., MLM) by injecting ``prompts'' and ``[MASK]'' to the PLM input. 
 Hence, the PLM can generate the desired results with minimal adjustment. This encourages downstream tasks to reuse the knowledge from the PLM \new{more efficiently}.

\section{Approach}
\label{sec:approach}

\subsection{Problem Definition and Analysis}
\noindent\new{\textbf{Definition 1} (\textit{Code Representation Learning}). 
Code representation learning aims at representing source code as vectors.}
Let $x=\{x_1,...,x_N\}\in\mathcal{X}$ denote a code snippet with $N$ tokens. A function $f$ is learned to map $x$ into a $d$-dimensional vector which contains the semantics of $x$~\citep{BengioCV13}, namely,
\begin{equation}
\label{eq:coderepresentation}
    f_\theta:\mathcal{X}\rightarrow R^d.
\end{equation}
$f_\theta$ is a function parameterized by $\theta$, which can be implemented using deep neural networks such as the fully-connected networks~\citep{code2vec}, LSTM~\citep{MahtoVTH21} and Transformers~\citep{transformer}. 

The learned program vectors can further be taken as input to machine learning models for code intelligence tasks, \new{such as code clone detection, code search, and method name prediction.}

\smallskip\noindent
\new{\textbf{Definition 2} (\textit{Code Classification Task}).}
Given two code or text fragments $x_1$ and $x_2$, a code classification task aims to predict a category ~$y\in\mathcal{Y}$ that represents their relationship: 
\begin{equation}
\label{eq:task}
    p(y|x_1, x_2) = g_\phi(f_\theta(x_1), f_\theta(x_2)),
\end{equation}
where $g_\phi$ denotes the neural classification model.
\new{Most of the code understanding tasks such as code clone detection, code search, and method name prediction can be formulated as a code classification task in a unified way.}
For example, in the code clone detection task, $x_1$ and $x_2$ stand for two code snippets and $y$ stands for whether they contain clones. In the code search task, $x_1$ and $x_2$ stand for a code snippet and a natural language description, respectively, and $y$ stands for whether they are semantically correlated.

\new{State-of-the-art code representation learning techniques usually leverage the \emph{pre-training and fine-tuning} paradigm: a Transformer is firstly pre-trained on large unlabeled code corpora using self-supervised objectives and is subsequently fine-tuned on labelled data of code classification tasks.}

\smallskip\noindent
\new{\textbf{Definition 3} (\textit{Zero-Shot Code Representation Learning}).}
\new{The goal of zero-shot code representation learning is to generate semantic representations of an unseen programming language (target language) without requiring the task-specific data. This can be achieved by reusing semantic representations of a seen programming language (source language).
Let $S$ be a source language and $T$ be the target language. }
\new{Zero-shot code representation learning aims to transfer the parameters $\theta$ from $f_\theta^S:\mathcal{X}^S\rightarrow R^d$ to $f_\theta^T:\mathcal{X}^T\rightarrow R^d$, where the former is trained on the task-specific source language training data.
}




\new{$f_\theta^S$ couldn't be used for the representation of $\mathcal{X}^T$ directly since there is a lexical gap between $\mathcal{X}^S$ and $\mathcal{X}^T$. However, both the source and the target language PLMs are pre-trained on large-scale unlabelled code corpora with self-supervised objectives such as the MLM for a PLM. Hence, it is feasible to bridge their representation gap by using the common knowledge learned in pre-training}. 
Based on this idea, we cast the problem in Equation~\ref{eq:task} as a pre-training task for PLM and train the model on ($\mathcal{X}^S$, $\mathcal{Y}^S$), so that the PLM can also \new{predict reasonable results for samples} in $\mathcal{X}^T$ seamlessly. 

\subsection{Model Architecture}
\begin{figure}[!t]
    \centerline{\includegraphics[width=\textwidth]{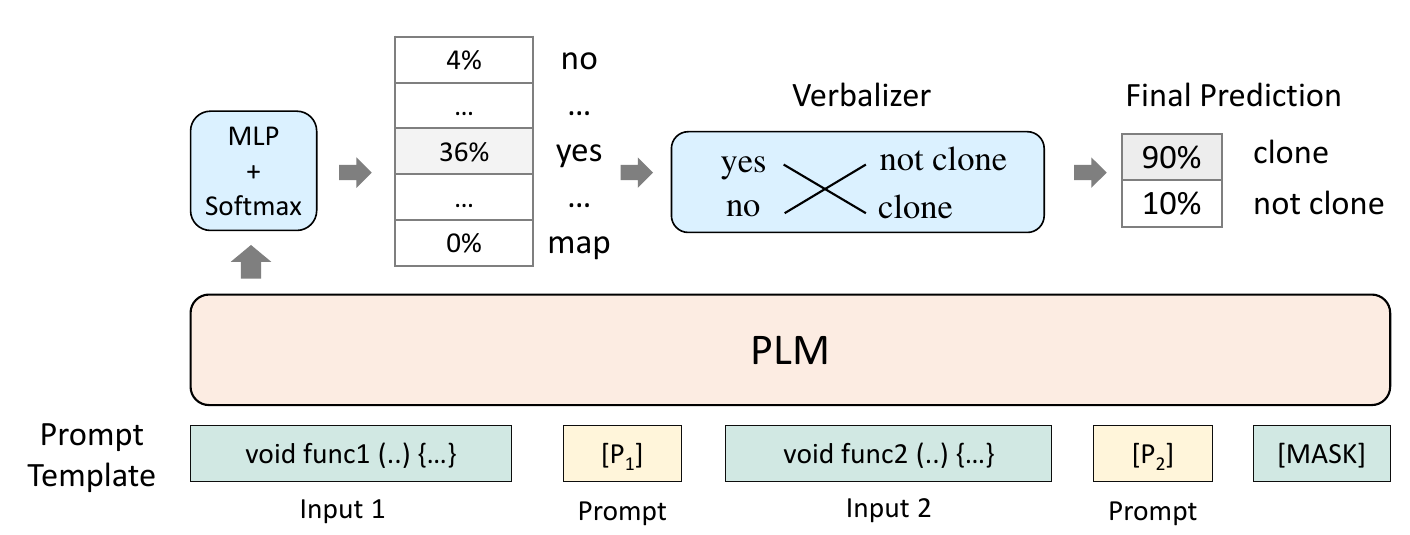}}
    \caption{\new{Model architecture of \approach for classification tasks.}}
    \label{fig:structure}
\end{figure}

Figure~\ref{fig:structure} illustrate the overall architecture of \approach.
The pipeline is comprised of three steps: 
\begin{itemize}
\item [1)] 
We first cast any downstream task to the pre-training task (e.g., MLM) by inserting trainable prompts and a ``[MASK]'' token into the input of the task. \new{The added prompts play the role of guiding the PLM to elicit the knowledge learned in pre-training and predict the right answer} (\S \ref{sec:cast}).
\item [2)] 
Taking the resulting data as input, a PLM is then \new{continually trained on the source language dataset of downstream tasks}, that is, infers the code representation of the input and predicts the word for the ``[MASK]'' token. \new{The optimal prompts are searched in the word embedding space automatically} (\S \ref{sec:prompt}). 
\item [3)] 
Finally, \new{we take unlabelled dataset in the target language as the input to the PLM and let the model predict the answer without training. A verbalizer is employed to cast the predicted word to class labels (\S \ref{sec:veb}).}
\end{itemize}


\subsection{Casting Downstream Tasks}
\label{sec:cast}
Our first step is to cast the downstream task (Equation~\ref{eq:task}) into the MLM task.
We concatenate two input snippets $x_1$ and $x_2$ of the downstream task,
\new{which can better capture the relationship between them~\citep{nspbert}. } 
Like that in the MLM task, we also insert an ``[MASK]'' token into the concatenated input. The ``[MASK]'' token acts as a placeholder which steers the pre-trained model to generate the classification result~$y$ in the code intelligence task. 
It is notable that the position of the ``[MASK]'' token is a hyperparameter and we append it at the tail of the input by default. 
The masked sequence 
\begin{equation}
    \tilde{x} = [CLS];x_1;x_2;[MASK]
\end{equation}
is taken as input to the PLM, which yields the hidden states
\begin{equation}
    \mathbf{h}_1,...,\mathbf{h}_{|\tilde{x}|} = f_\theta(\tilde{x})
\end{equation}
for all tokens. 
Then, the hidden state corresponding to the masked token, namely, $\mathbf{h}_{-1}$, is fed into an MLM header $g_\phi$ which predicts a token for the masked position: 
\begin{equation}
    \hat{\mathbf y} = \mathrm{softmax}(g_\phi(\mathbf{h}_{-1})).
\end{equation}
The MLM header $g_\phi$ is a fully connected neural network parameterized by $\phi$ that is optimized to minimize the cross-entropy loss:
\begin{equation}
\label{eq:mlmloss}
    L_{\mathrm MLM}(\phi|\mathbf{y}, \hat{\mathbf y}) = - \sum_{i=1}^{|V|}y_i\mathrm{log}(\hat{y}_i),
\end{equation}
where $\mathbf{y}$ denotes the ground-truth label of the code intelligence task, and $|V|$ is the vocabulary size.


\subsection{Prompt-based Learning}
\label{sec:prompt}
The conventional fine-tuning method optimizes $f_\theta$ and $g_\phi$ in Equation~\ref{eq:task} from scratch. This causes the model to overfit scarce task-specific data. Inspired by prompt-based learning~\citep{ptuning}, we optimize the PLM by merely adjusting its input sequence. More specifically, we insert a number of pseudo tokens called prompts into the input sequence of the PLM, which coax the PLM to directly generate the predicted class label of the downstream task. By only adjusting the model input, the PLM needs far less optimization cost to fit for the data in the target task, while keeping the most of prior knowledge learned during pre-training. 

Based on this idea, we design a number of prompt tokens~$P=[P_1,...,P_m]$ and inject them into the masked sequence~$\tilde{x}$ using a pre-defined template $T={[P]; x_1; [P]; x_2; [P]; [MASK]}$.
Hence, the original inputs $x_1$ and $x_2$ are transformed into
\begin{equation}
    \tilde{x} = [P_{1:i}]; x_1; [P_{i+1:j}]; x_2; [P_{j+1:m}];[MASK]
\end{equation}
through template $T$ which contains m prompt tokens.

Like general words, these prompt tokens are embedded into trainable vectors and are continually pre-trained on \new{downstream tasks of the source domain} through gradient descend optimization. 

\rev{However, the size of trainable parameters for prompts is too small compared with that of the original PLM. This may cause the prompt representation vectors to fall into local minima in gradient descent. To solve this problem, we add a bidirectional LSTM for encoding the prompt tokens. The LSTM encoder takes prompt tokens as input and outputs the hidden states. These hidden states, concatenated with the embeddings of the original code snippets, are fed into CodeBERT for the follow-up representations. } 

In a zero-shot setting, there is no training sample in \new{the target} low-resource language. Instead, we continually pre-train the PLM \new{and the trainable prompts} using large-scale code corpora in popular languages (e.g., Java), and then directly apply the trained model to tasks in the low-resource language. 
More specifically, \new{we train the PLM in the source domain through prompt-based learning}. Then, we take as input data samples in the target domain into the same model without extra training and obtain the results of the downstream tasks.

\subsection{Reverting Outputs to Final Answer}
\label{sec:veb}

The MLM task generates a token that is likely to fill the masked position. In order to obtain the classification result, we need to revert the MLM predictions to the classification labels of the downstream task. 
For this purpose, we employ a verbalizer~\citep{pet} which realizes such a reversion. Let $\mathcal{V}$ be the vocabulary of the PLM and $\mathcal{Y}$ be the labels of the downstream task such as $\{$true, false$\}$. The verbalizer is defined as a function $v$: $\mathcal{C} \rightarrow \mathcal{Y}$ that maps each candidate word in the vocabulary to a classification label. The choice of candidate words is arbitrary as long as they are sufficiently different. The model will be trained to map candidate words to true predictions. In our approach, we consider two candidate words $\{yes, no\}\in\mathcal{V}$ as a candidate set $\mathcal{C}$ and only inspect which word in $\mathcal{C}$ is more likely to fill into the ``[MASK]'' position through PLM predictions. If the word ``yes'' has a higher probability of filling in the masked position, the verbalizer will map it to the label ``true'' and hence output a positive prediction for this task.

Take code clone detection as an example. Given two code snippets, the model constructs an input sequence by injecting a number of prompt tokens into the snippets, followed by a ``[MASK]'' token. The constructed sequence is fed into the PLM to predict the label $Y$, where $Y\in\{$``cloned'', ``not cloned''$\}$. 
The MLM header of the PLM outputs the probability of each candidate word for the masked position. If the candidate word ``yes'' has a higher probability, the verbalizer will map it to the class label ``cloned'', yielding the final prediction as ``cloned''. 


\subsection{\approach for Generative Tasks}
\label{sec:generation}

\begin{figure}[!t]
    \centerline{\includegraphics[width=0.9\textwidth]{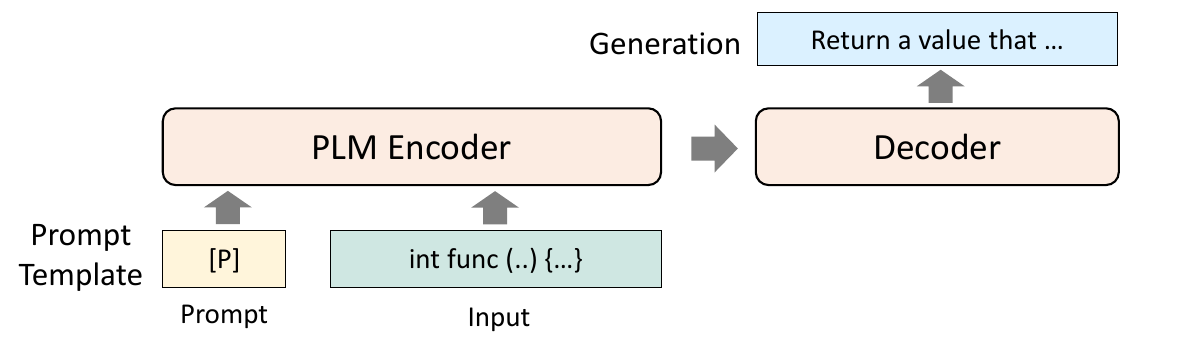}}
    \caption{\new{Model architecture of \approach for generative tasks.}}
    \label{fig:structure2}
\end{figure}

\new{Besides classification tasks, we also explore zero-shot learning on generative tasks such as program generation and code summarization.}

\smallskip\noindent
\new{\textbf{Definition 4} (\textit{Code Generative Task}).
A code generative task aims to generate the target sequence $y$ given an input code snippet or description $x$}: 

\begin{equation}
\label{eq:task2}
    p(y|x) = g_\phi(f_\theta(x)),
\end{equation}
\new{where $p$ denotes the probability of $y$, $f_\theta$ denotes the representation learner and $g_\phi$ denotes a neural generative decoder. 
For instance, the code summarization task generates a brief summary $y$ for the given code $x$.}

\new{Figure~\ref{fig:structure2} illustrates the architecture of \approach for generative tasks.}
\new{The model follows a conventional encoder-decoder architecture. The pipeline is comprised of two steps: First, the PLM encoder takes as input a code snippet or a description $x$, prepends it with a prompt $P$, and encodes $\tilde{x}$ as vectors. Then, the decoder generates the target sentence $y$ based on the encoded hidden vectors.}

\begin{equation}
    \tilde{x} = [P_{1:m}]; x;
\end{equation}

\new{The prompt $P$ is designed as a sequence of $m$ prefix tokens, which helps the PLM extract the knowledge learned in the pre-training phase.}
\new{In the code generation task, to specify the target language, we append a special token ``$<$language$>$'' to the original prompt where ``language'' indicates the target language name.}

\new{
We select CodeBERT as the encoder, and a randomly-initialed Transformer as the decoder.
Both the PLM and prompt are trained on the task-specific dataset in the source language and are then transferred to the generative task in the target language without extra training.}

\subsection{Training and Usage}
\begin{figure}[!tbp]
    \centerline{\includegraphics[width=0.9\textwidth]{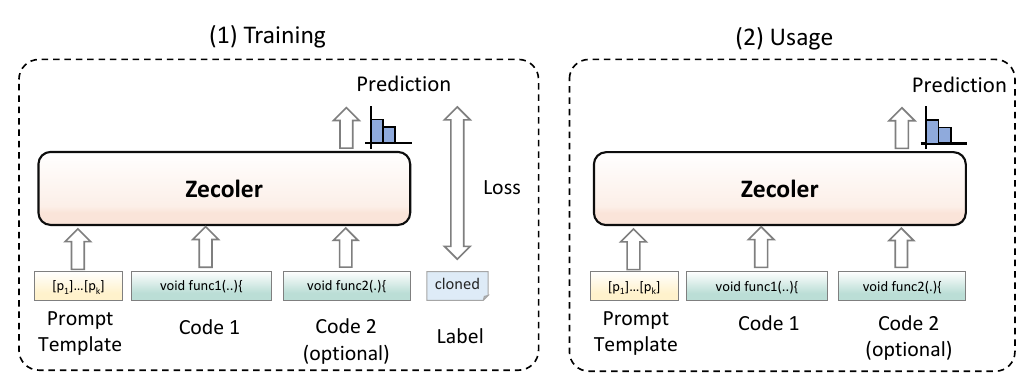}}
    \caption{The workflow of \approach. 
    }
    \label{fig:workflow}
\end{figure}

Figure~\ref{fig:workflow} shows the workflow of \approach. \approach follows the general paradigm of learning code representations. In the training phase, \approach is given a training set of labelled code snippets. For each snippet (pair), \approach augments it using a prompt template. The prompt-augmented code (pair) is taken as input to \approach which yields the prediction and calculates the loss function based on the ground-truth data. 

In the usage phase, \approach is given a code snippet (pair) only. \approach augments it using the same prompt template as in the training phase and then gives the prediction for the downstream task.

\section{Experimental Setup}
\label{sec:evaluation}

\subsection{Research Questions} 
We evaluate \approach by answering the following research questions:

    \begin{itemize}
        \item \textbf{RQ1: How effective is our approach in zero-shot code representation learning?}
        
        We evaluate the effectiveness of \approach in zero-shot code representation learning. We take Java as the source language and transfer the learned model to Solidity, a domain-specific languages, and Go, an up-and-coming language which are not provided with training samples. 
        The experiments are conducted in three popular classification tasks.
        
        \item \textbf{RQ2: How effective is our approach in few-shot code representation learning?}
        
        In some programming languages and tasks, we can obtain scarce (\eg, 100) samples in the target language. We wonder whether \approach is also effective in these data when pre-trained on the source domain. Therefore, we provide the pre-trained model in RQ1 with a few samples of the target language and conduct the same experiments as in RQ1.
        
        \item \textbf{RQ3: How effective is our approach in monolingual code representation learning?}
        
        RQ1 and RQ2 mainly evaluate the effectiveness of \approach in a cross-language setting. We further explore how effective is our approach without transfer learning. We train the model in three languages \new{in the few shot setting}, and test the model in the same language. Besides Solidity, we also want to assess our approach in other languages such as Java and Go when \new{training data is insufficient}. 
        
        \item \new{\textbf{RQ4: How effective is our approach in zero-shot generative tasks?}}
        
        \new{Besides the classification tasks, we also investigate the effectiveness of \approach in generative tasks. Similar to RQ1, we take Java as the source language and take Solidity and Go as the target language. We also take JavaScript and Ruby into account as the examples of program languages in the specific projects. The experiments are conducted in code summarization and code generation tasks.}
        
        \item \textbf{RQ5: How do different hyperparameters impact the performance of our approach?}
        
        We evaluate the performance of our approach under different hyperparameters. Specifically, we conduct ablation studies on prompt templates (number and position), source languages, and PLM scales.
        
        
        
    \end{itemize}

\subsection{Downstream Tasks} 
We evaluate our approach on \new{three classification tasks and two generative tasks}: 
\begin{itemize}
\item[1)]\textbf{Code Clone Detection (CD)}: a task that determines whether two code snippets are cloned or not~\citep{bigclonebench}. A PLM-based clone detection model takes as input two code snippets and outputs their representations. Then, a classification header is built on top of the representations and predicts whether the two code snippets are cloned (=1) or not (=0). There are four types of clones~\citep{clonedetection}. Our approach can challenge type-3 and type-4 clones, that is, the two snippets are not textually identical, but implement the same functionality.

\item[2)]\textbf{Code Search (CS)}: a task that retrieves a semantically relevant code snippet for a given natural language query~\citep{codesearchnet}. Following CodeBERT~\citep{codebert}, we formulate code search as a classification problem. Given a natural language description of the code and a programming language code snippet, this task aims at determining whether this NL-PL pair is related. The binary answer is ``related'' or ``not related''. The classification generates a probability score, which can be used for ranking results of code search.

\item[3)]\textbf{Method Name Prediction (MNP)}: a task that suggests the function name for a code snippet~\citep{ZhangCLP21}. Similar to code search, we transform this task to a binary classification task~\citep{ComptonFPK20}: given a code snippet, it enumerates all candidate function names (i.e., the vocabulary of code tokens) and constructs a ``$<$snippet, name$>$'' pair. The pair is taken as input to the PLM which outputs a binary prediction whether the name in the pair is ``suitable'' (=1) or ``not suitable''(=0) for the code snippet.

\item[4)]\new{\textbf{Code Summarization (CM)}: a task that generates a natural language summary for given source code~\citep{00030W0Z22}. We fine-tune the PLM using a parallel dataset of PL-NL pairs.}

\item[5)]\new{\textbf{Code Generation (CG)}: a task that automatically generates source code for a natural language query~\citep{WangWWMLZLWJL22}. Code generation is a challenging task because programs usually follow syntax rules while PLMs are good at generating sequential tokens. We fine-tune the PLM on a parallel set of NL-PL pairs.}

\end{itemize}

\subsection{\rev{Metrics}}
\rev{We measure the performance of all classification tasks (i.e., clone detection, code search, and method name prediction) using Accuracy, which is defined as the ratio of results that have been predicted correctly:}
\begin{equation}
    \mathrm{Accuracy} = \frac{\mathrm{\#\,of\,correct\,predictions}}{\mathrm{\#\,of\,all\,predictions}}
\end{equation}

\rev{We measure the performance of code summarization using BLEU and ROUGE-L, which are widely used metrics for generative tasks. 
BLEU measures the similarity of the machine-translated text to a set of high-quality reference translations. BLEU is computed as the ratio of predicted n-grams among all n-grams in the reference~\cite{bleu}. We compute the BLEU-4 score which refers to the average BLEU score when n=1, 2, 3, and 4. The ROUGE-L metric~\cite{lin-2004-rouge} is also a widely used metric for summarization, which computes the score based on the length of the longest common subsequence present in the reference text and the hypothesis text.}

\rev{We measure the performance of code generation using CodeBLEU and ROUGE-L. CodeBLEU is an extension of BLEU for source code~\cite{codebleu}. Besides the text n-grams, CodeBLEU takes into account the match of syntactic units in abstract syntax trees.}

\subsection{Datasets}

\begin{table}[!t]  
	\centering
	\caption{Overview of datasets.}
	\label{tab:dataset}
	\resizebox{\textwidth}{!}{
	\begin{tabular}{lccccccc}  
		\toprule		      \multirow{2}{*}{Datasets} & \multicolumn{5}{c}{Downstream Tasks *} & \multirow{2}{*}{Size} & \multirow{2}{*}{Programming Languages} \\ \cline{2-6}
		\quad & CD & CS & MNP & \new{CM} & \new{CG} &  &   \\  
		\midrule   
		SCCD & $\surd$ & & & & & 10,000 & Solidity \\
		SCS & & $\surd$ & $\surd$ & $\surd$ & $\surd$ & 347,410 & Solidity \\
		CodeNet & $\surd$ & $\surd$ & & & & 8,008,527 & Java, Go, C++, C, Python, Ruby, C\#, ...\\
		CodeSearchNet & & & $\surd$ & $\surd$ & $\surd$ & 2,000,000 & Java, Go, Python, JavaScript, Ruby, PHP \\
		\bottomrule  
	\end{tabular}}
\begin{tablenotes}
\item[*] CD = clone detection, CS = code search, MNP = method name prediction, \new{CM = code summarization, CG = code generation}.
\end{tablenotes}	
\end{table}

We conduct experiments on four datasets. Each dataset may be used for multiple downstream tasks. Table~\ref{tab:dataset} shows the statistics of each dataset, including sizes, programming languages, and corresponding tasks.

\smallskip\noindent\textbf{Smart Contract Clone Detection (SCCD)}: a manually labeled clone detection dataset for the Solidity language. The dataset contains 10,000 data samples that are collected from EtherScan\footnote{https://etherscan.io/}, an analytic platform for smart contracts. We build a web scraper to collect Solidity code and label the cloned pairs based on contract information such as contract address and opcode. Each data sample consists of a pair of code snippets that are cloned. \rev{More specifically, we download solidity files from EtherScan and extract subcontracts from each file. We randomly select a pair of subcontracts, label them as a clone pair, and randomly sample an equal number of false clone pairs from different files. To reduce noise, two postgraduate students from the research lab manually labeled these contracts and resolved conflicts through discussions until a consensus was reached.} One notable feature of this dataset is that most samples are type-3 and type-4 clones.

\smallskip\noindent\textbf{Smart Contract Summarization (SCS)}~\citep{YangKYGWMZ21}: a dataset that contains 347,410 code-comment pairs in the Solidity language. The dataset was originally collected for code summarization, and we preprocess it to fit for the code search and method name prediction tasks. To adapt to the code search task, we filter long code and remove code comments. For method name prediction, we separate method names from original code snippets. 

\smallskip\noindent\textbf{CodeNet}~\citep{codenet}: a multilingual codebase built from two online judge websites, namely, AIZU\footnote{https://onlinejudge.u-aizu.ac.jp/introduction} and AtCoder\footnote{https://atcoder.jp/}. CodeNet contains 8,008,527 code submissions in multiple programming languages such as Java, Go, Ruby, and Python. We use this dataset for the code clone detection and code search tasks. To adapt the dataset to the code clone detection task, we label two code submissions as a cloned pair if they solve the same problem. 
To adapt the original data to the code search task, we extract (NL, PL) pairs from problem descriptions and their code submissions, respectively. 

 
\smallskip\noindent\textbf{CodeSearchNet}~\citep{codesearchnet}: a widely used dataset for NL-PL and PL-NL tasks. The dataset involves 2,000,000 code snippets in six languages, namely, Java, Go, Python, JavaScript, Ruby and PHP. Each snippet is accompanied with \new{a corresponding natural language description and the} method name.

We preprocess these datasets by removing comments \new{from the code since they can interfere with the final results in classification task experiments.} \rev{Since the PLM takes the concatenation of two code snippets as input, we restrict each code snippet to have a maximum of 250 tokens. We also exclude short code (i.e., less than 125 tokens) that occurs rarely in real scenarios of downstream tasks. This process helps unify the lengths of input code and improves the model performance in downstream tasks.}
In order to prevent the model from being biased to one class in classification tasks, we balance the dataset with the same number (1:1) of positive and negative samples. 
The negative pairs ($y=0$) are created using random combinations of snippets from the positive data samples ($y=1$). 

\subsection{Implementation Details}

We implement all models on top of CodeBERT, a popular PLM which is built based on RoBERTa-base (H=768, A=12, L=12). CodeBERT learns representations of programming languages (Java, Python, JavaScript, PHP, Ruby, and Go) in the pre-training phase. We use the default tokenizer (\ie, Microsoft/codebert-base) of CodeBERT with a vocabulary size of 50,265. We set the maximum sequence length to 512. 
Our experimental implementation is based on the Huggingface Transformers\footnote{https://huggingface.co/microsoft/codebert-base} and P-Tuning~\citep{ptuning}. \rev{All hyperparameters were tuned to the optimal values through extensive tuning on the validation set.} The batch size and the number of epochs are set to 10 and 20  \new{in classification tasks, and 20 and 15 in generative tasks.} 

In classification tasks, we insert prompt tokens uniformly into the original input of CodeBERT \new{since there are two text snippets in the input. }
\new{The additional LSTM for training prompts has two hidden layers followed by two-layer multilayer perceptrons (MLP) activated by ReLU.}

\new{To generate target sequences, we employ a 6-layer Transformer decoder as a fine-tuning header.}
\new{Since we append a special prompt ``$<$language$>$'' to the original input and change the input pattern, we continually pre-train CodeBERT using MLM task on 100,000 unlabelled code sampled from CodeSearchNet, each appended with a language mark. The batch size and number of epochs are set to 8 and 3, respectively.}

All models are optimized using the AdamW~\citep{adamw} algorithm on a machine with a GeForce RTX 3090 Ti GPU \rev{(24G) and a main memory of 32GB}. The initial learning rate (lr) is set to 3e-5, which linearly increases from 0 during a warm-up period. The iteration number of the warm-up period equals to the number of the training steps in the first epoch. During the rest training process, the learning rate continuously decreases to 0.
We measure the performance on the validation set during training. The checkpoint that achieves the best accuracy on the validation set is selected for testing. 

\subsection{Baseline Models}
\label{sec:baseline}

We compare our approach with five baseline models: 
\begin{itemize}
\item[1)] \textbf{AVG}: a baseline approach that directly represents programs by averaging their token embeddings. We reuse token embeddings from CodeBERT and represent an input code snippet by the average of all token embeddings. Next, we fine-tune the classifier of downstream tasks using a 3-layer MLP header. 


\item[2)] \textbf{RoBERTa}~\citep{roberta}: a popular pre-trained language model that has also been used for programming languages~\citep{codebert}. The model is constructed with 12 transformer layers and pre-trained on a large English corpus with the MLM objective. We fine-tune it with a 3-layer MLP header over the ``[CLS]'' position.

\item[3)] \textbf{RoBERTa-large\footnote{https://huggingface.co/roberta-large}}:  
a large version of RoBERTa (H=1024, A=16, L=24)  with around 300 million parameters. We compare with this model to verify the advantages of \approach over large-scale PLMs.

\item[4)] \textbf{CodeBERTa\footnote{https://huggingface.co/huggingface/CodeBERTa-small-v1}}:
a version of RoBERTa pre-trained with CodeSearchNet, which was proposed by Huggingface. We use its default setting in our experiments.

\item[5)] \textbf{CodeBERT}~\citep{codebert}: one of the state-of-the-art models for learning code representations. A more detailed description of CodeBERT can be found in Section~\ref{sec:pretrain}. We follow the same experimental setup in its original paper.

\end{itemize}

We implement these baseline models by referring to the work of CodeXGlue~\citep{codexglue}. In classification tasks, we construct 3-layer fully connected neural networks as the fine-tuning header which maps the hidden vector of the ``[CLS]'' token to the class labels of downstream tasks. \new{In generative tasks, we use a 6-layer transformer decoder as the fine-tuning header for generating the target sequence of downstream tasks.}

\rev{We did not compare \approach with stronger code PLMs such as Codex and CodeT5. Codex and CodeT5 have been shown to be more effective in generative tasks such as code generation, while the primary goal of Zecoler is code representation, namely, representing source code into vectors. Therefore, CodeBERT is a more appropriate backbone model regarding the problem setting and architecture design. Furthermore, our approach relies on p-tuning, which is built on an encoder-only architecture and has been shown to be more effective for representation tasks.}

\section{Results}

\subsection{RQ1: Effectiveness of Zero-shot Learning}
In this experiment, we evaluate the effectiveness of \approach in zero-shot code representation learning. 
We initially train a representation model for each task using data samples of Java. \rev{Java is one of the most popular programming languages. It has a large availability of code data}. Then, we adapt the trained model to the target languages (i.e., Solidity and Go) directly without extra training. We train the model with both 5,000 and 500 data samples of Java to assess the effects under different data sizes.

\begin{table}[!t]
	\centering
	\caption{Accuracy of code representation models on three classification tasks in the zero-shot setting.}
	\label{tab:rq1}
	\begin{threeparttable}
    \setlength{\tabcolsep}{3.1mm}{
	\begin{tabular}{l@{}ccccccccccc}
	\toprule
    \multirow{2}{*}{Model} &
    \multirow{2}{*}{\makecell[c]{Number of \\ Layers}} &
    \multicolumn{2}{c}{CD} & \multicolumn{2}{c}{CS} & \multicolumn{2}{c}{MNP}\\ \cline{3-8} 
                              & &Solidity&Go&Solidity&Go&Solidity&Go\\ \hline
    AVG       &   -      &   57.5    & 49.2      &  49.0     & 50.3     &  50.8     &  50.0  \\                          
    RoBERTa  &  12                  &   60.5    & 49.4      &  49.6     & 49.4      &  50.0     &  50.3  \\
    RoBERTa-L &  24       &   47.3    & 51.0       & 48.7      & 48.8       &  51.7     &  48.5  \\
    CodeBERTa  &  6        &   57.9    &  67.3      &   53.2    & 53.1       &  49.7  & 49.0    \\
    CodeBERT  &  12                 &   65.4    & 91.7       &  48.9     & 46.2       &  52.1     &  65.2  \\\hline
    Zecoler\tiny{ 5000}       &  12 & \textbf{79.8}    & \textbf{96.4}       &  \textbf{67.1}     & \textbf{80.3}        &  59.2     &  \textbf{98.8}   \\
    Zecoler\tiny{ 500}   & 12        &   74.9    & 82.4      &  53.3     & 56.9      &  \textbf{68.1}     &  90.4   \\ 
    \bottomrule
    \end{tabular}}
\begin{tablenotes}
\item[*] The target languages (i.e., Solidity and Go) are not provided with training data. All source languages are trained with 5000 samples except the last one which is trained with only 500 samples.
\end{tablenotes}    
\end{threeparttable}
\end{table}

Table~\ref{tab:rq1} shows the accuracy of different models in three classification tasks. We can observe that \approach significantly outperforms baseline models in all three tasks and all target languages. In the code clone detection task, the accuracy of \approach is 5\%-14\% greater than that of CodeBERT, the strongest baseline. The improvement is much more significant in the code search (30\% in average) and method name prediction (24\% in average) tasks. By contrast, AVG and RoBERTa-large obtain results that are close to random, indicating that they can hardly learn useful knowledge from few data samples.


The same trend can be observed when only 500 (1/10) samples of the source language are provided for training.
As the data size decreases from 5000 to 500, the accuracy of \approach drops in all tasks. Nevertheless, it still significantly outperforms the baseline models. 
This means that \approach can learn representations much more efficiently while requiring smaller data compared with baselines.

\new{Another interesting observation is that \approach trained with 500 data samples outperforms that with 5,000 data samples in the method name prediction task for Solidity. One potential reason is that the model overfits the big Java data in the training process, which results in inferior performance in the target low-resource languages such as Solidity.}

It is notable that CodeBERT outperforms RoBERTa and CodeBERTa in both code clone detection and method name prediction, except for the code search task. We conjecture that CodeBERT is pre-trained on programming languages whereas the RoBERTa is only pre-trained on natural languages. Hence, CodeBERT can be better adapted to PL related tasks.





\smallskip\noindent\textbf{Answer to RQ1:}
Our approach shows greater performance than baseline models in no-resource code classification tasks, affirming the strong ability of \approach in zero-shot code representation learning.


\subsection{RQ2: Effectiveness of Few-Shot Learning}

In this experiment, we evaluate the effectiveness of \approach in few-shot learning of code representations. \new{This simulates the scenario where there are a few training data samples in the target language.}
We continue training the model in RQ1 using a few data samples of the target languages. We vary the data sizes from 32 to 700 \new{in Solidity and Go} and evaluate the performance in three classification tasks.

\begin{figure}[tb]
    \centering
    \subfigure[Clone Detection (Solidity)]{
        \includegraphics[scale = 0.15, trim=10 10 10 10]{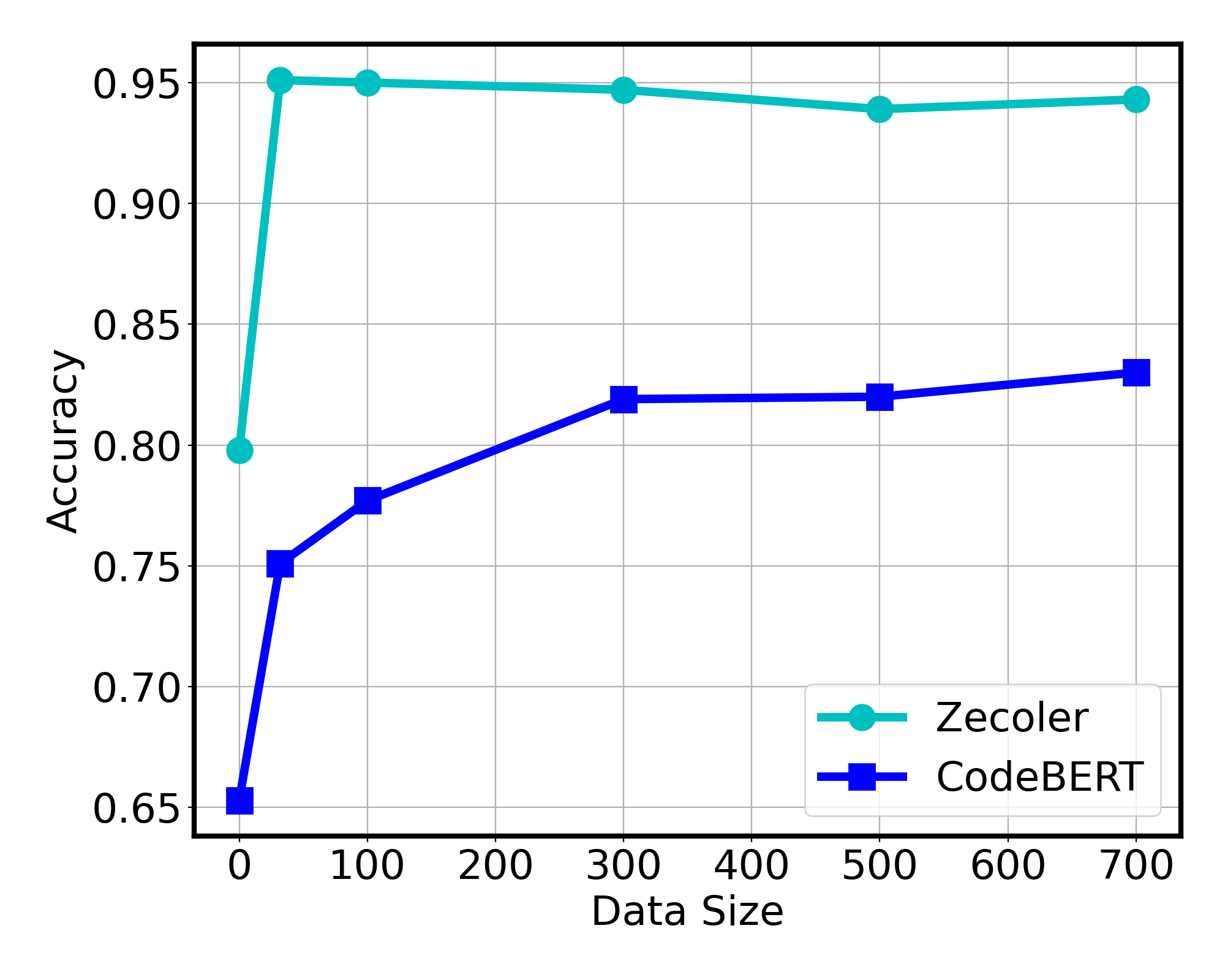}
    }
    \subfigure[Code Search (Solidity)]{
        \includegraphics[scale = 0.15, trim=10 10 10 10]{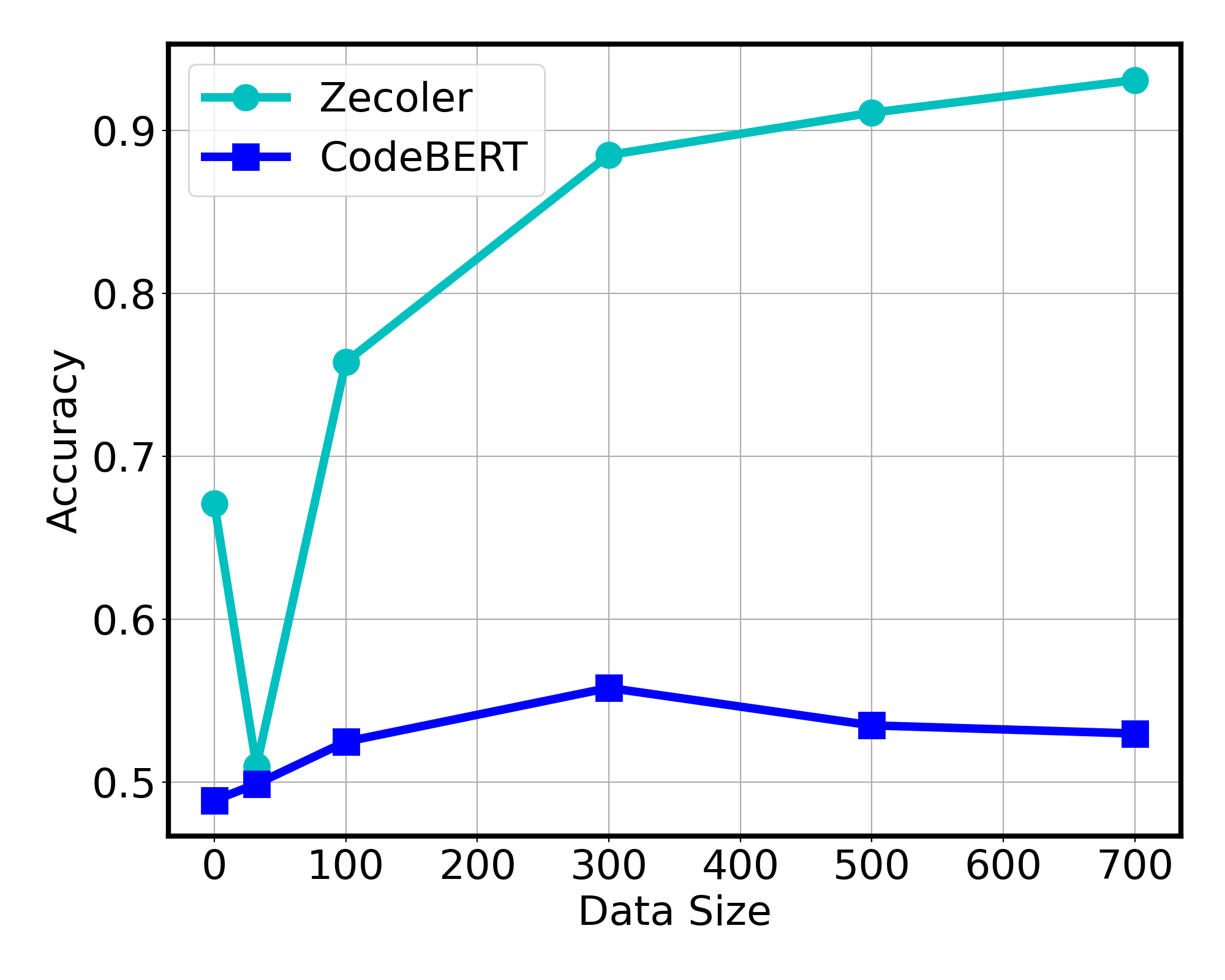}
    }
    \subfigure[Method Name Prediction (Solidity)]{
        \includegraphics[scale = 0.15, trim=10 10 10 10]{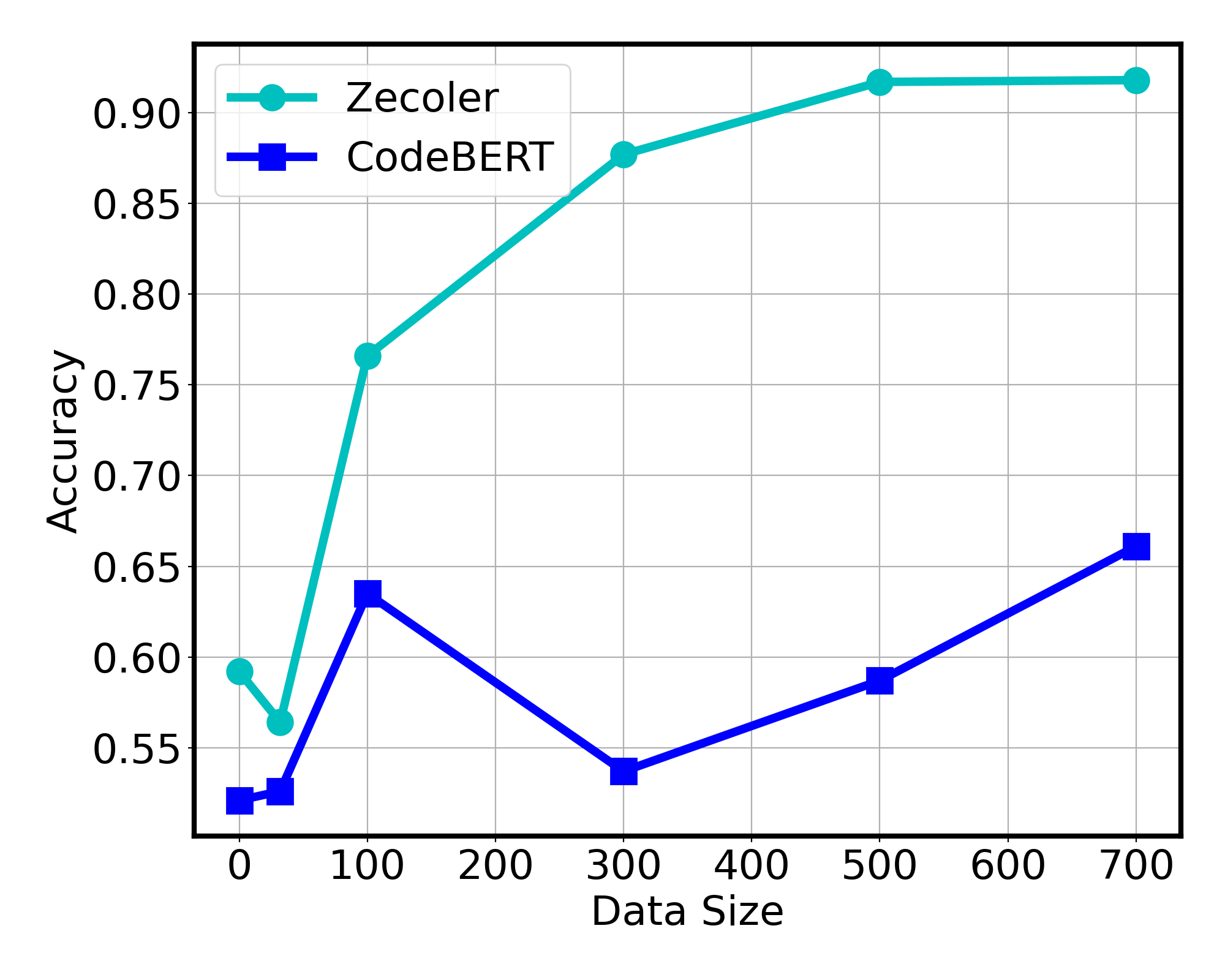}
    } \\
    \vspace{-6pt}

    \subfigure[Clone Detection (Go)]{
        \includegraphics[scale = 0.15, trim=10 10 10 10]{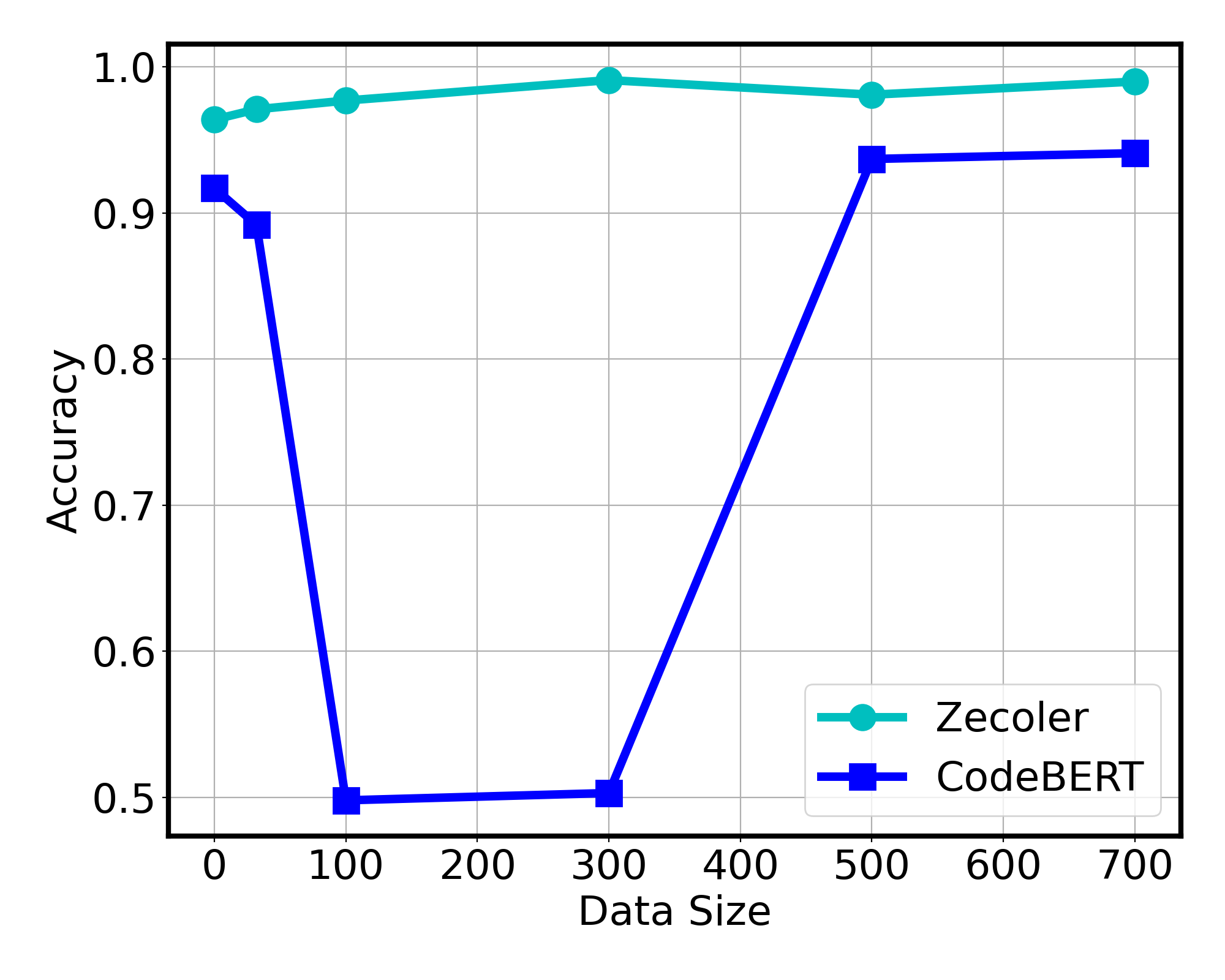}
    }
    \subfigure[Code Search (Go)]{
        \includegraphics[scale = 0.15, trim=10 10 10 10]{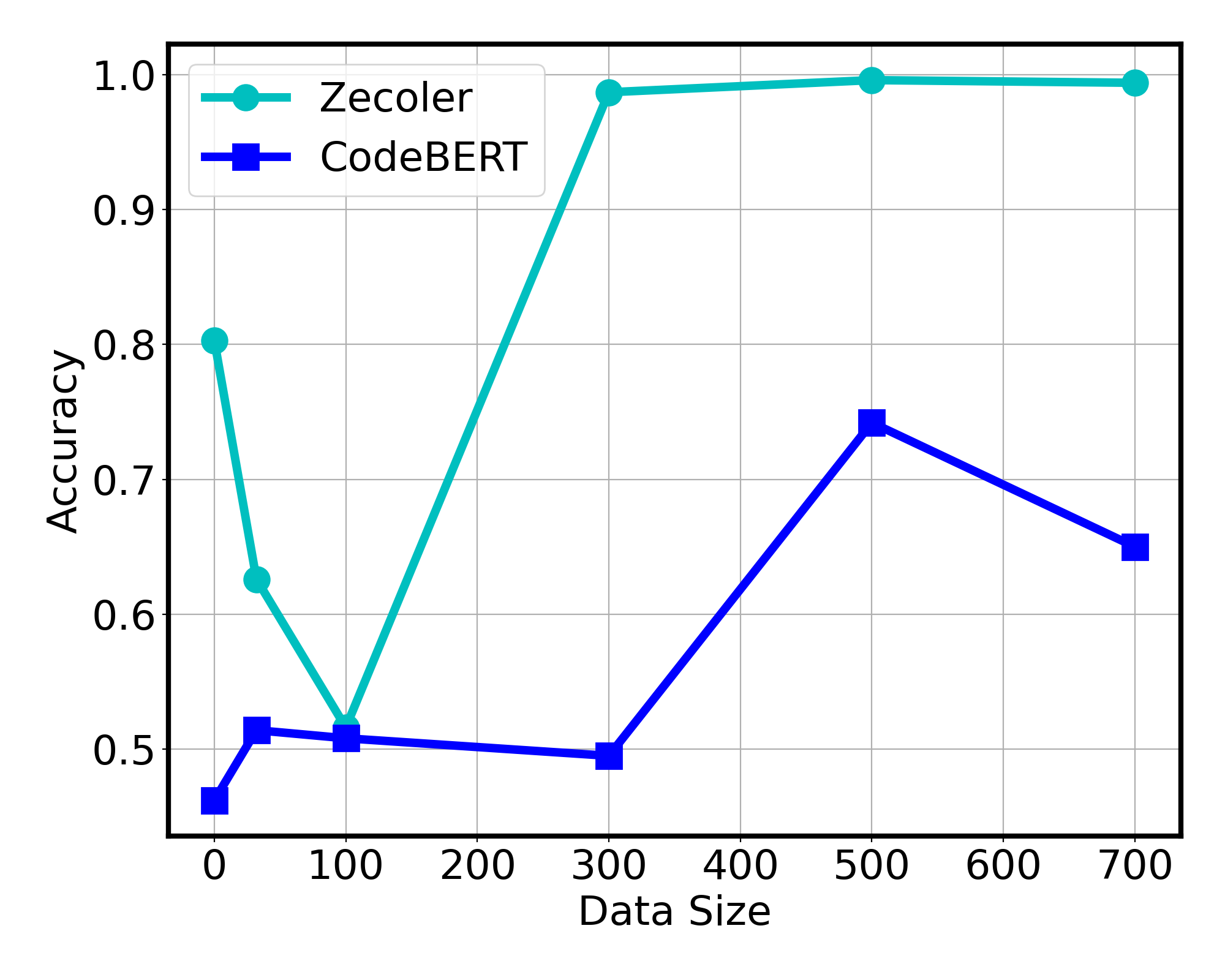}
    }
        \subfigure[Method Name Prediction (Go)]{
        \includegraphics[scale = 0.15, trim=10 10 10 10]{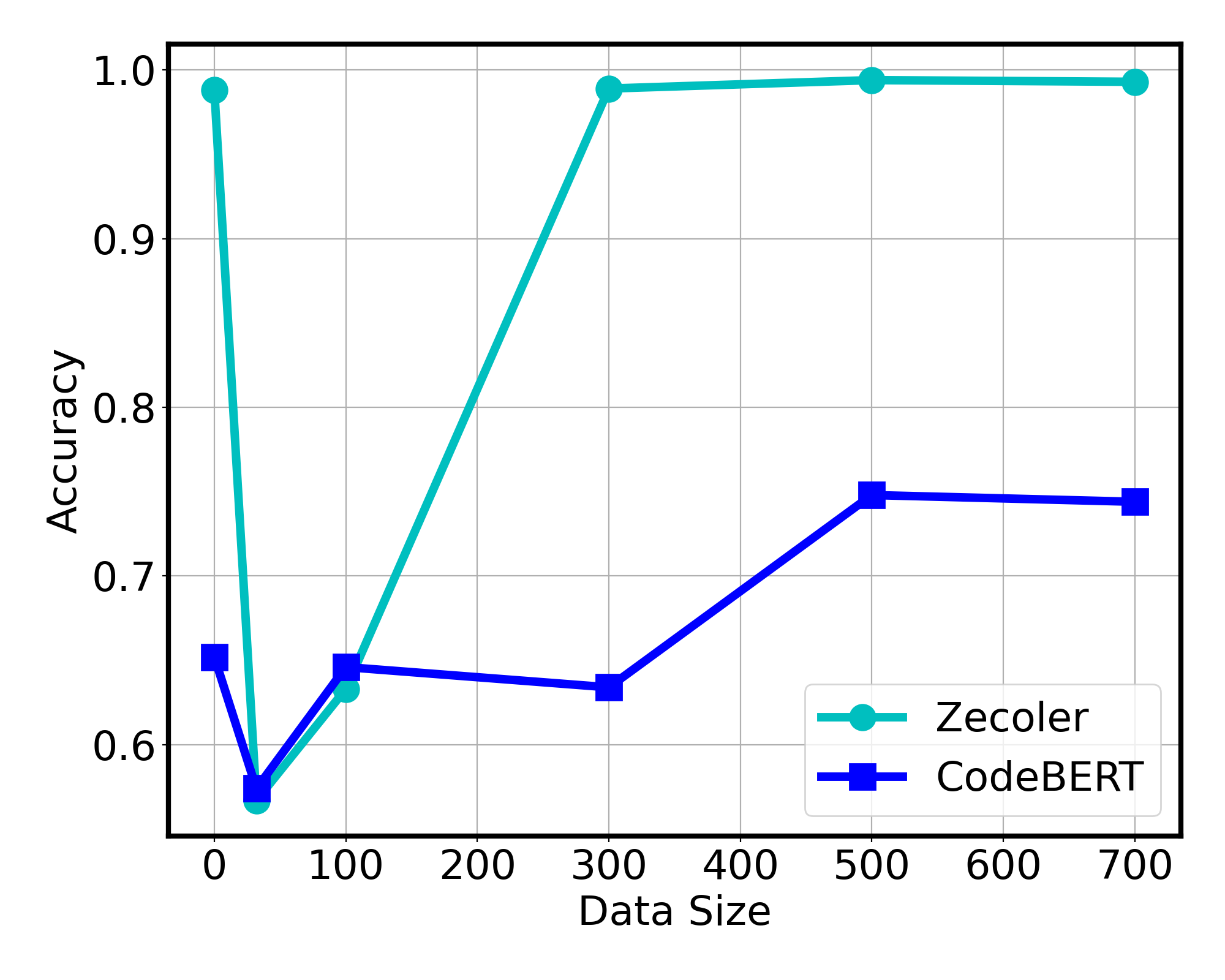}
    }\\

    \caption{Accuracy of code representation models continuously trained with different-scale datasets in few-shot setting in classification tasks.
    }
    \label{fig:rq2}
\end{figure}

Figure~\ref{fig:rq2} shows the results in classification tasks. Compared to CodeBERT, the strongest baseline in RQ1, \approach demonstrates greater strengths in all tasks when provided with a few data samples of the target language. When the data size is 700, the accuracy of \approach is about 30\% greater than that of CodeBERT. 
This means that \approach outperforms existing approaches in learning code representations even when a small number of samples are given. \new{In RQ1, the model has already learned knowledge of downstream tasks in the training phase. It is easier to achieve good performance by continually training on the target domain instead of training from scratch.}

We notice that when the data size is extremely small (e.g., 32), the model tends to overfit data. In this situation, zero-shot learning is preferred.

\smallskip\noindent\textbf{Answer to RQ2:}
\approach demonstrates effective performance when a few data samples of the target domain are given. 
As in the zero-shot setting, \approach still keeps the best accuracy or BLEU among all compared models in few-shot setting. 

\subsection{RQ3: Effectiveness of Monolingual Few-Shot Learning}

Different from RQ2 in a cross-language few-shot setting, in this experiment we evaluate the effectiveness of our approach in a monolingual few-shot setting. We train the models with a few samples of Java, Solidity, and Go, and evaluate the performance of the tasks in the same language. 

\begin{table*}[!t]  
	\centering
	\caption{Accuracy of code representation models \new{on classification tasks} in a monolingual few-shot setting.}
	\label{tab:rq3}
	\resizebox{\textwidth}{!}{
	\begin{threeparttable}
	\begin{tabular}{lcccccccccc}
	\toprule
    \multirow{2}{*}{Model} & \multicolumn{3}{c}{CD} & \multicolumn{3}{c}{CS} & \multicolumn{3}{c}{MNP} & \multirow{2}{*}{Average} \\ \cline{2-10} 
                              & Java     & Solidity & Go     & Java   & Solidity  & Go     & Java    & Solidity  & Go   &   \\ \hline
    AVG\tiny{ 300}               &  51.6    & 64.1     & 50.1   & 49.1   & 50.2      & 50.1   & 47.9    &  48.1     & 50.1 & 51.3  \\
    RoBERTa\tiny{ 300}                    &  46.6    & 73.5     & 50.1   & 52.5   & 55.2      & 50.1   & 50.1    &  53.7     & 50.1 & 53.5  \\
    RoBERTa-large\tiny{ 300}              &  53.0    & 75.9     & 53.9   & 50.4   & 57.4      & 51.4   & 47.8    &  55.6     & 50.1 & 55.1 \\
    CodeBERTa\tiny{ 300}    & 50.7    & 68.3     & 65.3  &  52.0  & 58.8     & 45.0   & 50.8  &  61.8     & 47.4  & 55.6 \\
    CodeBERT\tiny{ 300}                   &  51.3    & 69.4     & 49.5   & 50.8   & 56.5      & 49.5   & 49.3    &  53.9     & 49.5 & 53.3  \\\hline
    Zecoler\tiny{ 300}        &  \textbf{85.8}    & \textbf{94.3}     & 99.3   & 51.7   & \textbf{90.1}      & \textbf{99.5}   & \textbf{98.7}    &  \textbf{88.8}     & \textbf{99.2} & \textbf{89.7}  \\
    Zecoler\tiny{ 100}        &  63.6    & 93.9     & \textbf{99.5}   & \textbf{52.6}   & 63.0      & 95.7   & 72.8    &  62.8     & 77.7  & 75.7 \\ 
    \bottomrule
    \end{tabular}
\begin{tablenotes}
\item * In this experiment, we train and test the model in the same programming language respectively. The training data size of source languages is 300 except for the last one with only 100 data samples.
\end{tablenotes}    
\end{threeparttable}
}
\end{table*}

Table~\ref{tab:rq3} shows the accuracy of different approaches in three classification tasks. 
We can observe that \approach consistently outperforms baselines in the monolingual few-shot setting. 
Most of the baseline models simply predict random answers, with an accuracy of around 50\%. This indicates that the baseline models cannot learn meaningful code representations with scarce data. Comparatively, \approach achieves 75.7\% accuracy in average with only 100 data samples. The results suggest that \approach learns code representations efficiently in the monolingual few-shot setting. 
\rev{We also noticed that \approach archives poor performance on Java code search. This might be due to the variance within the 100 samples. We also conjecture that there might be overfitting since the model was prompt-tuned with only 100 samples.}

\begin{figure}[tb]
    \centering
    \subfigure[Clone Detection (Solidity)]{
        \includegraphics[scale = 0.15, trim=10 10 10 10]{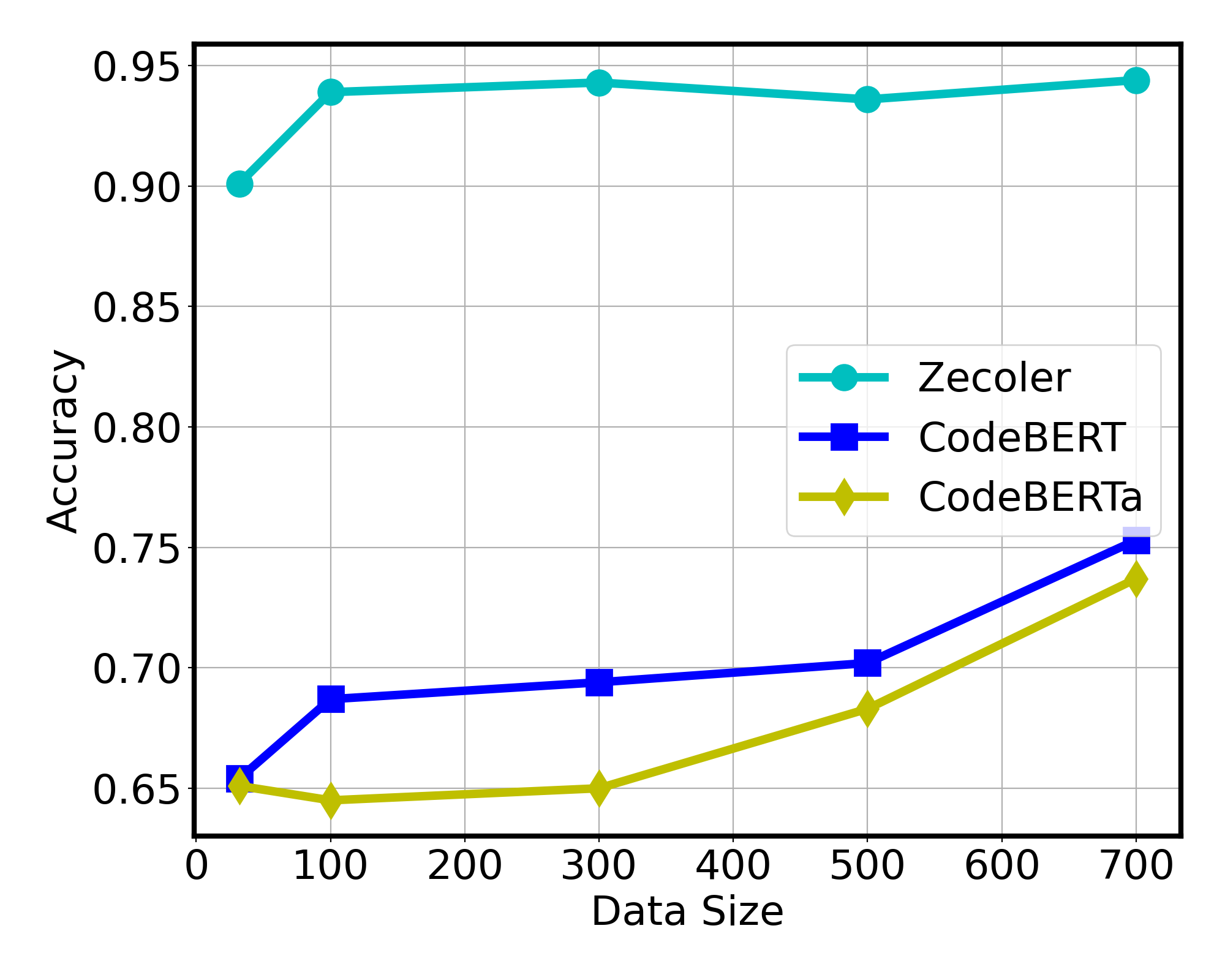}
    }
    \subfigure[Code Search (Solidity)]{
        \includegraphics[scale = 0.15, trim=10 10 10 10]{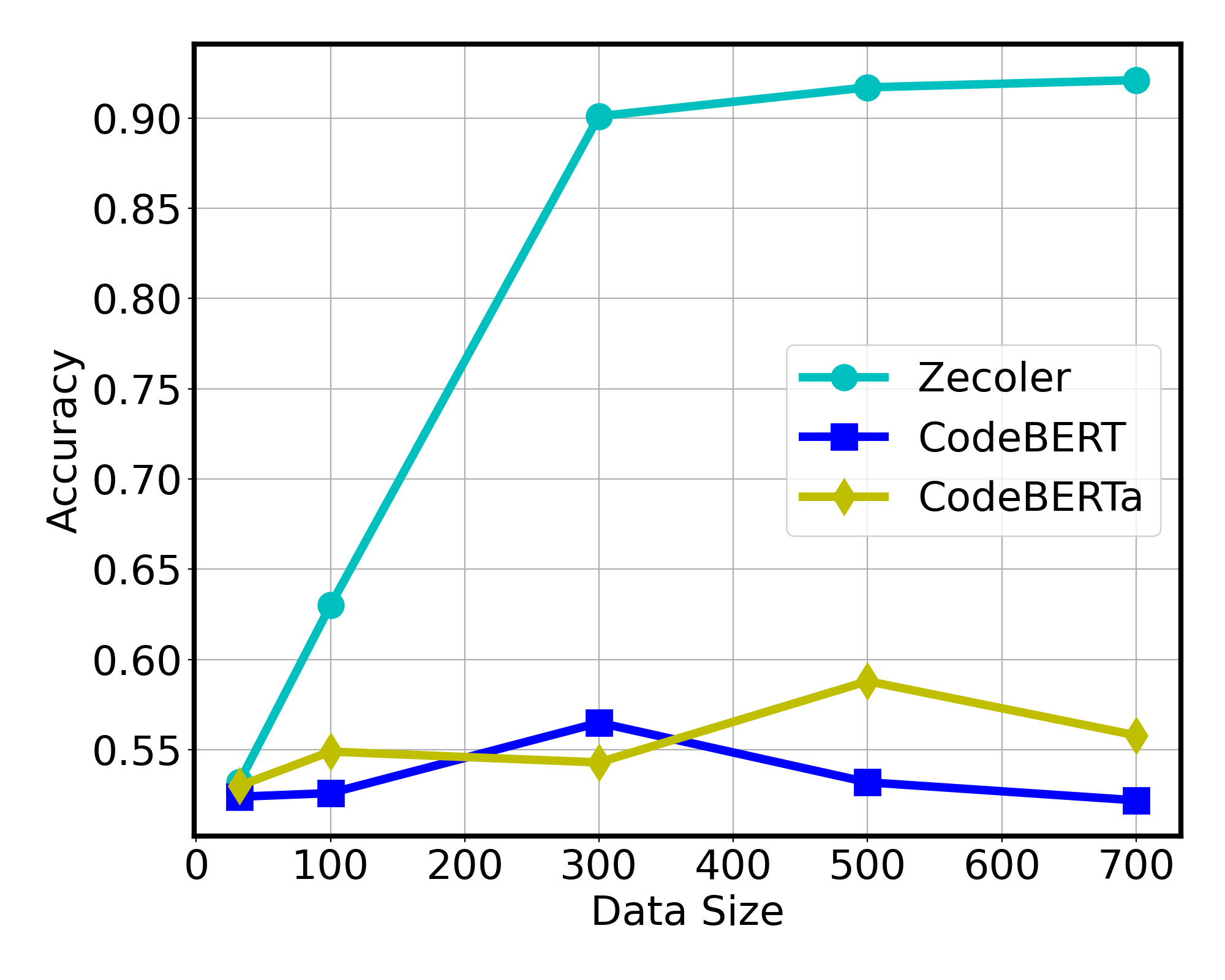}
    }
    \subfigure[Method Name Prediction (Solidity)]{
        \includegraphics[scale = 0.15, trim=10 10 10 10]{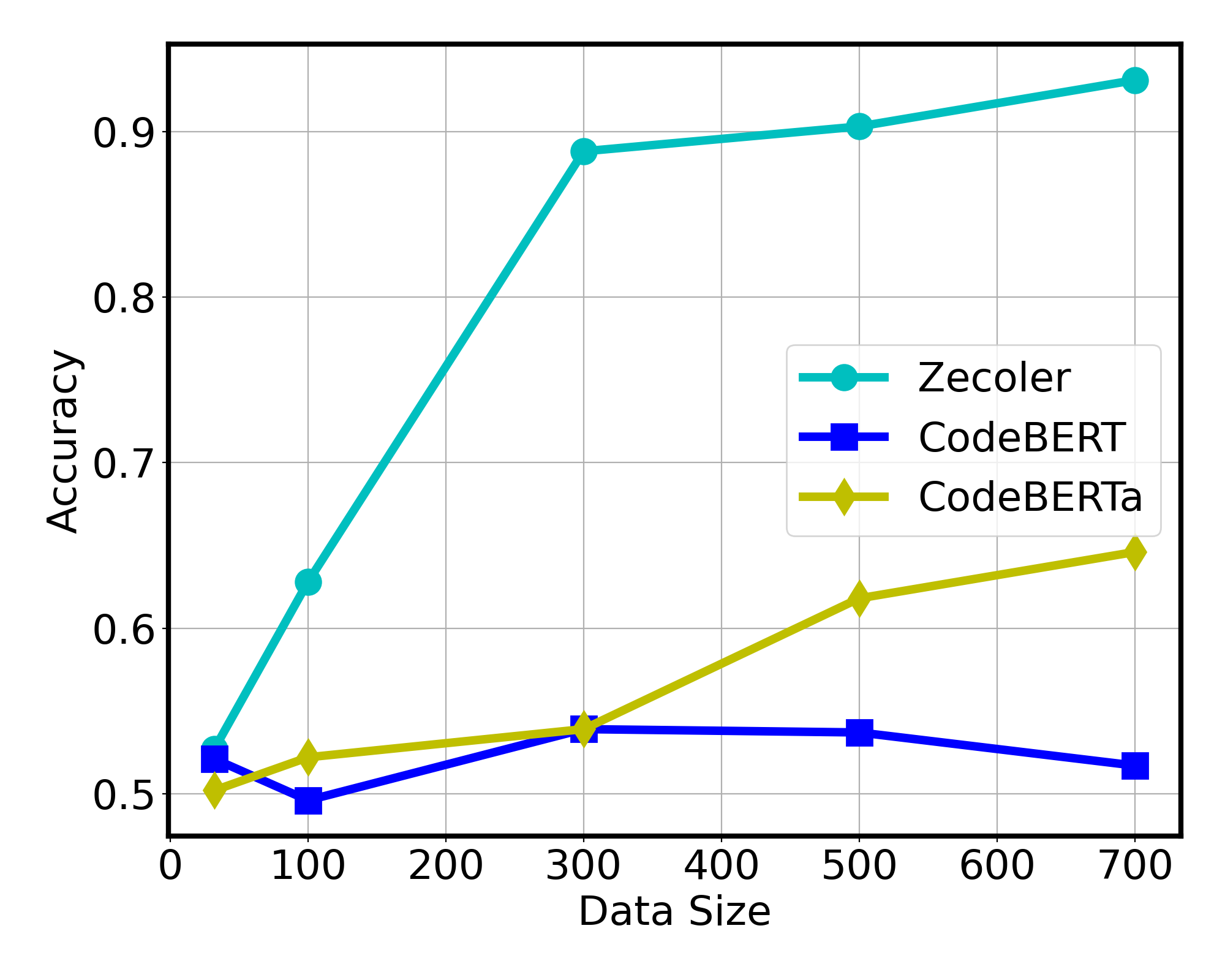}
    }\\    
    \vspace{-6pt}
    \subfigure[Clone Detection (Go)]{
        \includegraphics[scale = 0.15, trim=10 10 10 10]{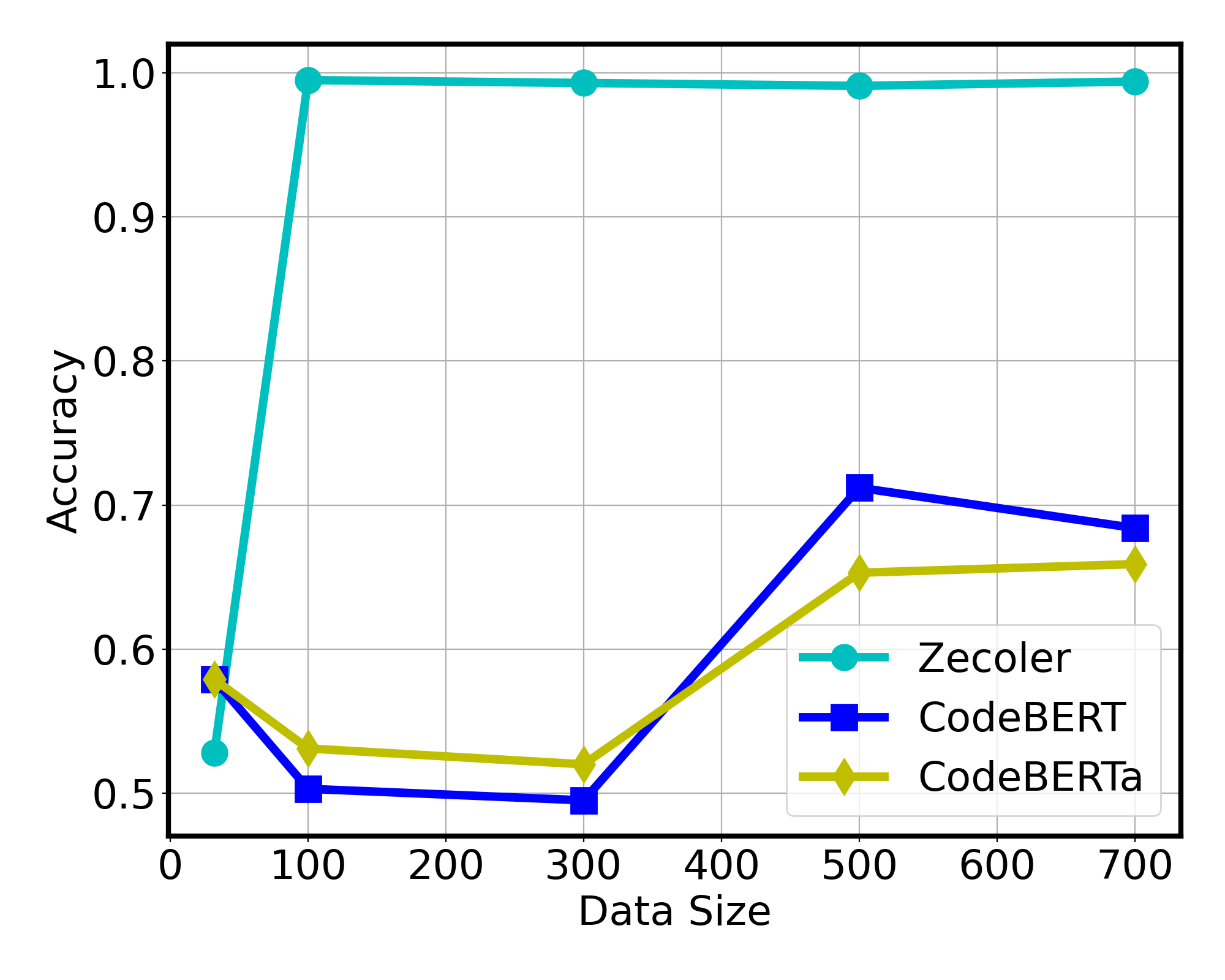}
    }
    \subfigure[Code Search (Go)]{
        \includegraphics[scale = 0.15, trim=10 10 10 10]{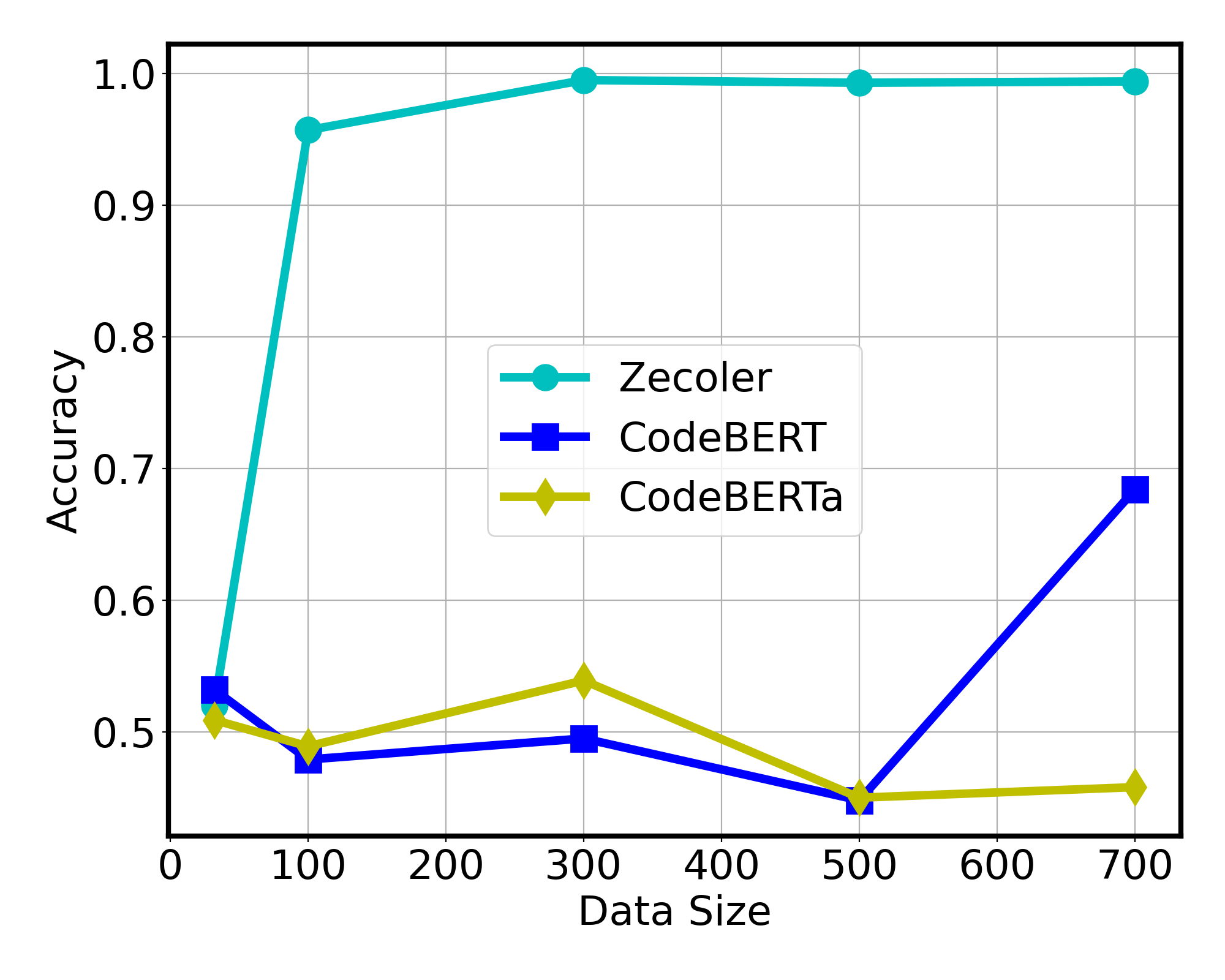}
    }
    \subfigure[Method Name Prediction (Go)]{
        \includegraphics[scale = 0.15, trim=10 10 10 10]{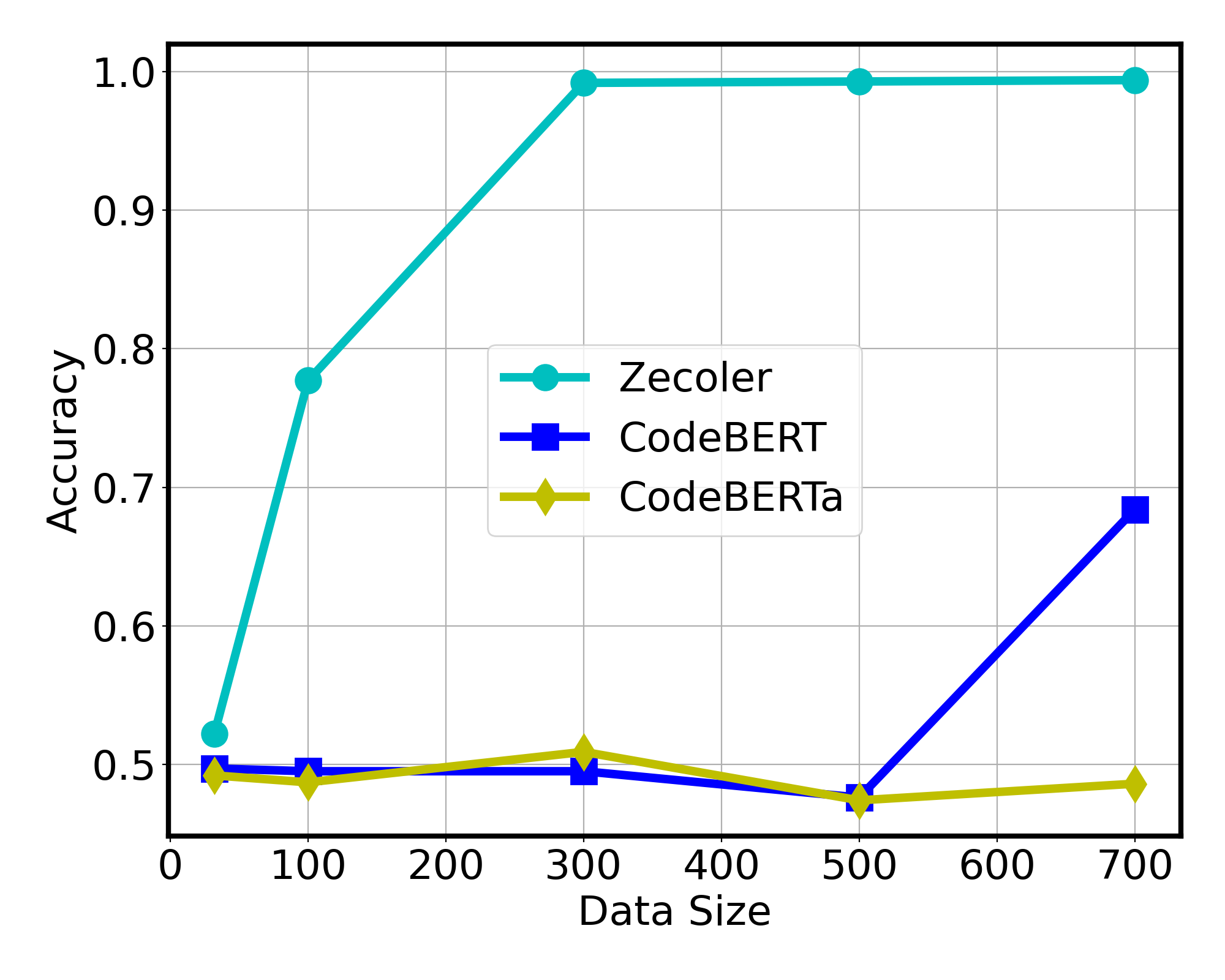}
    }\\    
    \caption{Accuracy of code representation models continuously trained with different-scale datasets in monolingual few-shot setting in classification tasks.
    }
    \label{fig:rq3}
\end{figure}

Figure~\ref{fig:rq3} shows the performance of \approach, CodeBERT, and CodeBERTa with different data sizes in the classification task. 
We can see that \approach outperforms the other two baselines under almost all data sizes. Furthermore, as the data size increases, the accuracy of \approach grows faster than that of baseline models. This indicates that \approach is effective in learning code representations given only a few data samples. 

We have also observed that monolingual learning outperforms cross-language learning on small data sizes (\eg, 32 and 100), but achieves similar performance when the data size becomes larger. This is because continual training on scarce data of a different language can lead to overfitting. 






\smallskip\noindent\textbf{Answer to RQ3:}
\approach is effective in monolingual few-shot learning, and demonstrates much stronger performance than that in the cross-language setting.

\subsection{RQ4: Effectiveness of Zero-shot Learning in Generative Tasks}

\new{In this experiment, we evaluate the effectiveness of \approach in zero-shot generative tasks.
Similar to the experiment in RQ1, we initially train a representation model for each task using 5,000 data samples of Java, and then apply the trained model to the target languages (i.e., Solidity and Go) without extra training. }

\begin{table}[!t]
\centering
\caption{\new{Accuracy of zero-shot code representation on code summarization task.}}
\label{tab:rq1cm}
\resizebox{\textwidth}{!}{
\new{
\begin{tabular}{l@{}cccccccc}
\toprule
\multirow{2}{*}{Model} & \multicolumn{2}{c}{Solidity} & \multicolumn{2}{c}{Go} & \multicolumn{2}{c}{JavaScript} & \multicolumn{2}{c}{Ruby} \\ \cline{2-9} 
                       & BLEU        & ROUGE-L        & BLEU     & ROUGE-L     & BLEU         & ROUGE-L         & BLEU      & ROUGE-L   \\ \hline
RoBERTa                & 11.35       & 19.12          & 7.96     & 15.55       & 6.78         & 11.79           & 7.71      & 14.24               \\
RoBERTa-L              & 12.52       & 20.30          & 8.43     & 16.61       & 8.01         & 13.84           & 9.07      & 16.10               \\
CodeBERTa              & 12.01       & 17.69          & \textbf{8.74}     & 16.77       & 7.75         & 12.05           & 8.90      & 15.84              \\
CodeBERT               & 12.68       & 19.39          & 8.21     & 17.64       & 8.60         & 15.40           & 9.45      & 17.05              \\ \hline
Zecoler                & \textbf{13.37}       & \textbf{20.67}          & 8.67     & \textbf{17.73}       & \textbf{9.54}         & \textbf{16.88}           & \textbf{9.97}      & \textbf{17.59}        \\      
\bottomrule
\end{tabular}}}
\end{table}

\begin{table}[!t]
\centering
\caption{\new{Accuracy of zero-shot code representation learning on code generation task.}}
\label{tab:rq1cg}
\resizebox{\textwidth}{!}{
\new{
\begin{tabular}{l@{}cccccccccc}
\toprule
\multirow{2}{*}{Model} & \multicolumn{2}{c}{Solidity} & \multicolumn{2}{c}{Go} & \multicolumn{2}{c}{JavaScript} & \multicolumn{2}{c}{Ruby} \\ \cline{2-9} 
                       & BLEU      & ROUGE-L      & CodeBLEU   & ROUGE-L   & CodeBLEU       & ROUGE-L       & CodeBLEU    & ROUGE-L    \\ \hline
RoBERTa                & 2.69             & 19.09        & 7.88       & 17.68     & 7.55           & 13.71         & 9.05        & 19.43      \\
RoBERTa-L              & \textbf{4.01}             & 20.29        & \textbf{9.55}       & 18.09     & 8.51           & 14.87         & 10.02       & 20.69      \\
CodeBERTa              & 3.29             & 20.12        & 7.70       & 18.42     & 7.11           & 13.61         & 9.14        & 20.47      \\
CodeBERT               & 2.74             & 19.29        & 8.68       & 18.13     & 8.36           & 15.01         & 9.67        & 20.51      \\ \hline
Zecoler                & 3.20             & \textbf{20.78}        & 9.18       & \textbf{18.78}     & \textbf{8.52}           & \textbf{15.26}         & \textbf{10.05}       & \textbf{21.26} \\      
\bottomrule
\end{tabular}}}
\end{table}

\new{Table~\ref{tab:rq1cm} and \ref{tab:rq1cg} show the accuracy of various models in code summarization and code generation tasks, respectively.} 
\new{We measure the accuracy of code generation using CodeBLEU~\citep{codebleu} instead of BLEU except in Solidity which is not supported by CodeBLEU.}
\new{We can observe that \approach consistently outperforms baseline models across most target languages in two tasks. Compared with CodeBERT,  
\approach gains 6.68\% and 4.84\% improvement in code summarization and 4.35\% and 4.33\% improvement in code generation in terms of BLEU and ROUGE-L.}

\rev{We also notice that RoBERTa-large and CodeBERTa surpass our approach in the code summarization task of Go. The main reason can be that these models have a bigger size (i.e., more trainable parameters), which will be further discussed in RQ5.}

\smallskip\noindent\new{\textbf{Answer to RQ4:}
Our approach consistently outperforms existing approaches in zero-shot generative tasks, indicating good generalizability of \approach.} 

\subsection{RQ5: Ablation Study}
\label{sec:ablation}
In this experiment, we inspect the performance of \approach under different hyperparameters. We vary the prompt templates and numbers of prompt tokens to search for the optimal prompt template. We also explore the impact to the performance by different source languages and different scales of backbone PLMs.

\smallskip\noindent\textbf{Prompt Templates:}
We first explore the effect of prompt templates on the performance. 
We vary the position of the prompt tokens $P_{1:k}$ in the prompt template, namely,
head: [$P_{1:k}$, $x_1$, $x_2$, MASK], 
middle: [$x_1$, $P_{1:k}$, $x_2$, MASK], uniformly: [$P_{1:m}$, $x_1$,$P_{m+1:n}$, $x_2$,$P_{n+1:k}$, MASK] and
tail: [$x_1$, $x_2$, MASK, $P_{1:k}$].
The number of prompt tokens ($k$) is fixed to 10.
We train the model with 700 Java code snippets and evaluate the model on the code clone detection task of Solidity. 

\begin{figure}[tb]
    \centering
    \subfigure[\scriptsize{Ablation on the position of prompts}]{
        \includegraphics[scale = 0.15, trim=-150 10 -150 10]{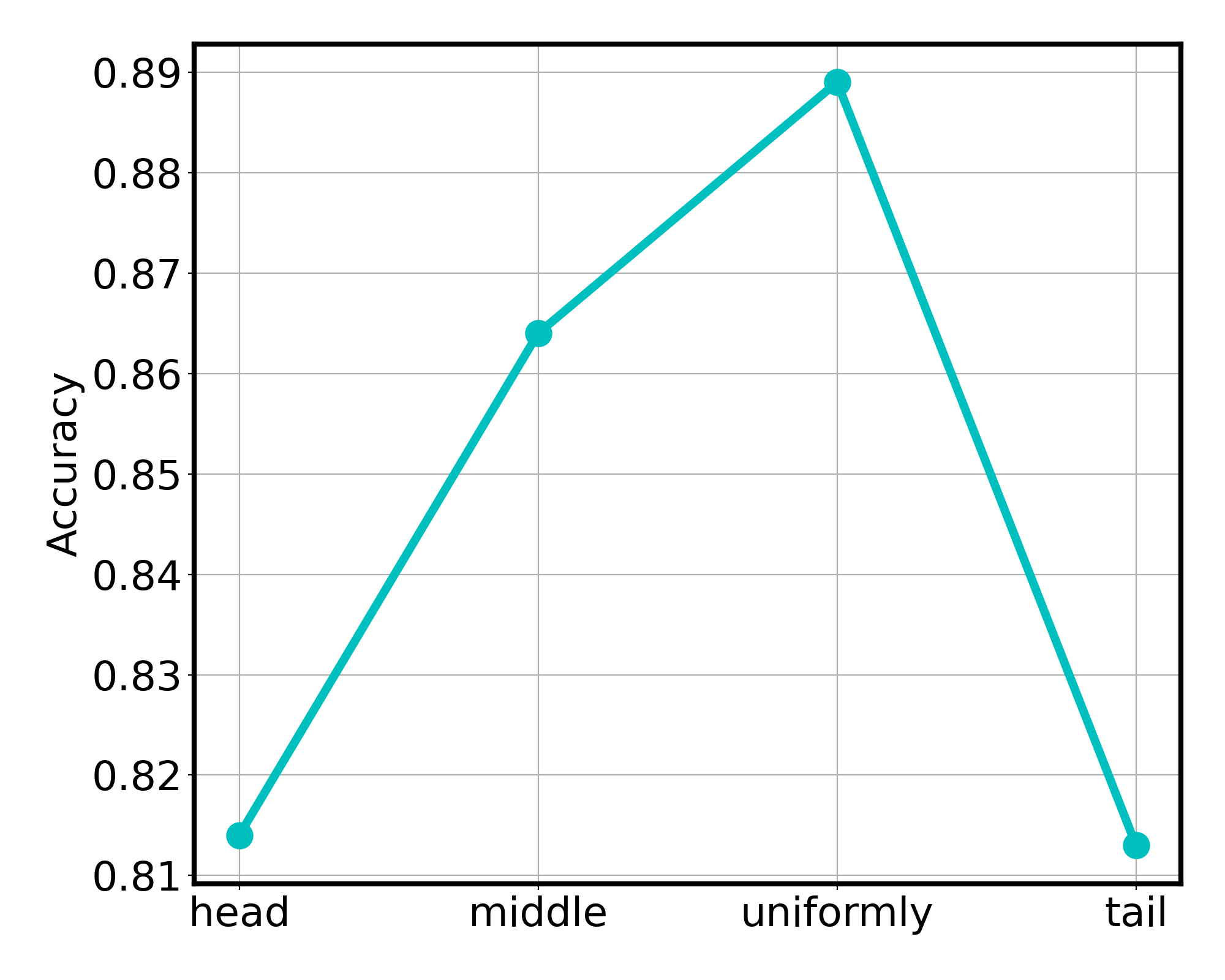}
        \label{fig:promptpos}
    }    
    \subfigure[\scriptsize{Ablation on the number of prompts}]{
        \includegraphics[scale = 0.15, trim=-150 10 -150 10]{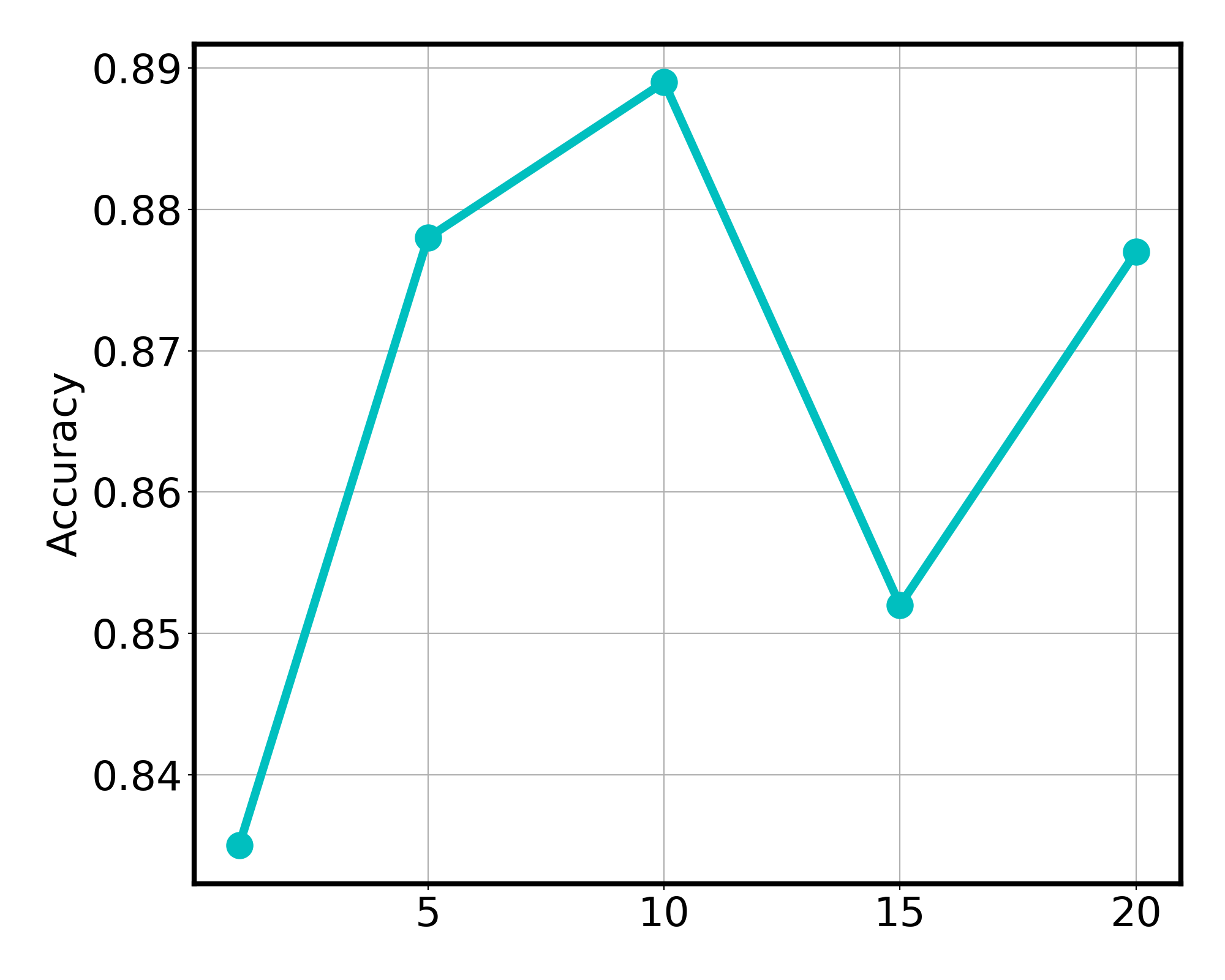}
        \label{fig:promptnum}
    }

    \caption{Performance of \approach under different prompt positions and prompt numbers on the Solidity clone detection dataset (SCCD).}
\end{figure}

As shown in Figure~\ref{fig:promptpos}, placing prompt tokens uniformly achieves the best performance compared to other templates. The reason could be that prompts have more influence to nearby tokens. By placing prompts uniformly, every input token can be influenced by sufficient prompts. 

\smallskip\noindent\textbf{Number of Prompt Tokens:}
We further assess the impact of prompt numbers. We insert prompt tokens uniformly into the PLM input and vary the number of prompt tokens~$k$ from 1 to 20. We train the model with 700 Java code snippets and evaluate it on the code clone detection task of Solidity.

As Figure~\ref{fig:promptnum} shows, the number of prompt tokens is strongly correlated to the performance of representation learning. Fewer prompt tokens are insufficient to steer the PLM to yield meaningful prediction, while large numbers of prompts restrict the input size. The optimal number of prompt tokens is 10 in our experiments.
\new{A similar test has been made in generative tasks and the optimal number of prompt tokens is also 10.}

\begin{figure}[tb]
\centerline{\includegraphics[width=0.8\textwidth]{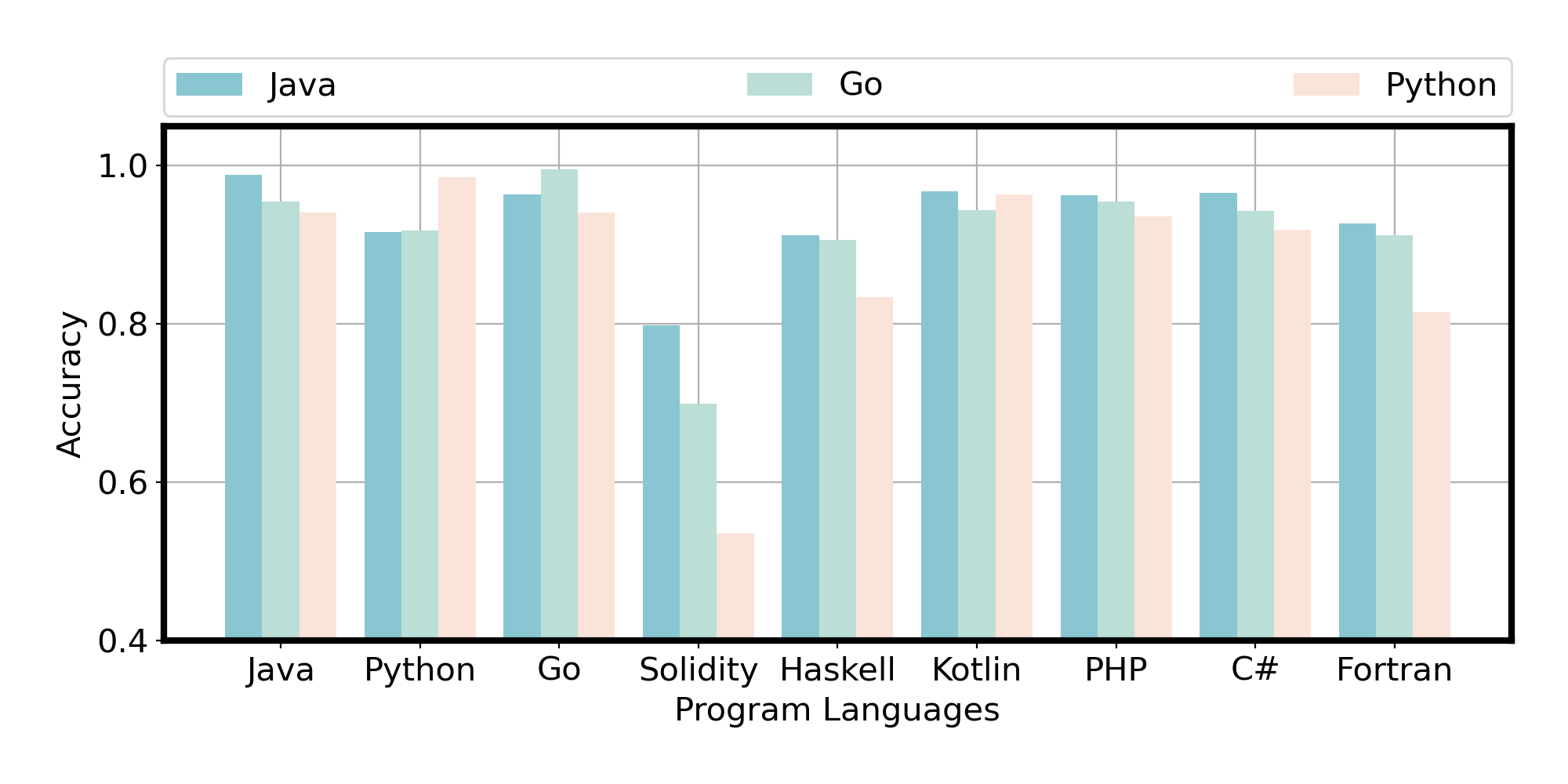}}
\caption{Performance of \approach in the code clone detection task with different source languages.}
\label{fig:diffsource}
\end{figure}

\smallskip\noindent\textbf{Source Languages:} 
To study the impact of different source languages, we train the model using 5,000 data samples of Java, Python, and \rev{Go}, respectively. We evaluate the performance of zero-shot code representation on the code clone detection task of nine target languages in the CodeNet.
Figure~\ref{fig:diffsource} shows the results. We can observe that using Java as the source language achieves the best performance. This can be attributed to \rev{the high popularity of Java code in the pre-trained corpus of CodeBERT}. 

\smallskip\noindent\textbf{Pre-trained model size:} 
Lastly, we study the effect of model size. In classification tasks (Table~\ref{tab:rq1} and \ref{tab:rq3}), larger PLMs have a negative effect on the performance. For example, RoBERTa-large is almost twice the size of RoBERTa, but the former performs even worse than the latter. \new{On the contrary, in generative tasks (Table~\ref{tab:rq1cm} and \ref{tab:rq1cg}), larger PLMs perform better than those in classification tasks.}  

\new{This discrepancy could be caused by different model structures between classification and generative tasks. In classification tasks, the models need to encode two separate \rev{code snippets which is different from the pre-training objective where only one code snippet is taken as input. Hence, classification tasks are more sensitive to model sizes}. By contrast, PLMs for generative tasks take a single code snippet as input. Hence, larger PLMs, with more common knowledge learned in the pre-training phase, have more potential to enhance the downstream tasks. 
}

\smallskip\noindent\textbf{Answer to RQ5:}
The effectiveness of our approach is affected by prompt templates, source languages, and PLM scales. Inserting ten prompt tokens uniformly to the original PLM input can better steer the PLM to learn code representations. Java as the source language can be better generalized to other languages. \new{In the zero- or few-shot setting, PLM scale does not always make a positive impact, which is closely related to the type of task.}

	
\subsection{\new{Qualitative Analysis}}

\begin{figure}[!t]
    \centerline{\includegraphics[width=1.0\textwidth]{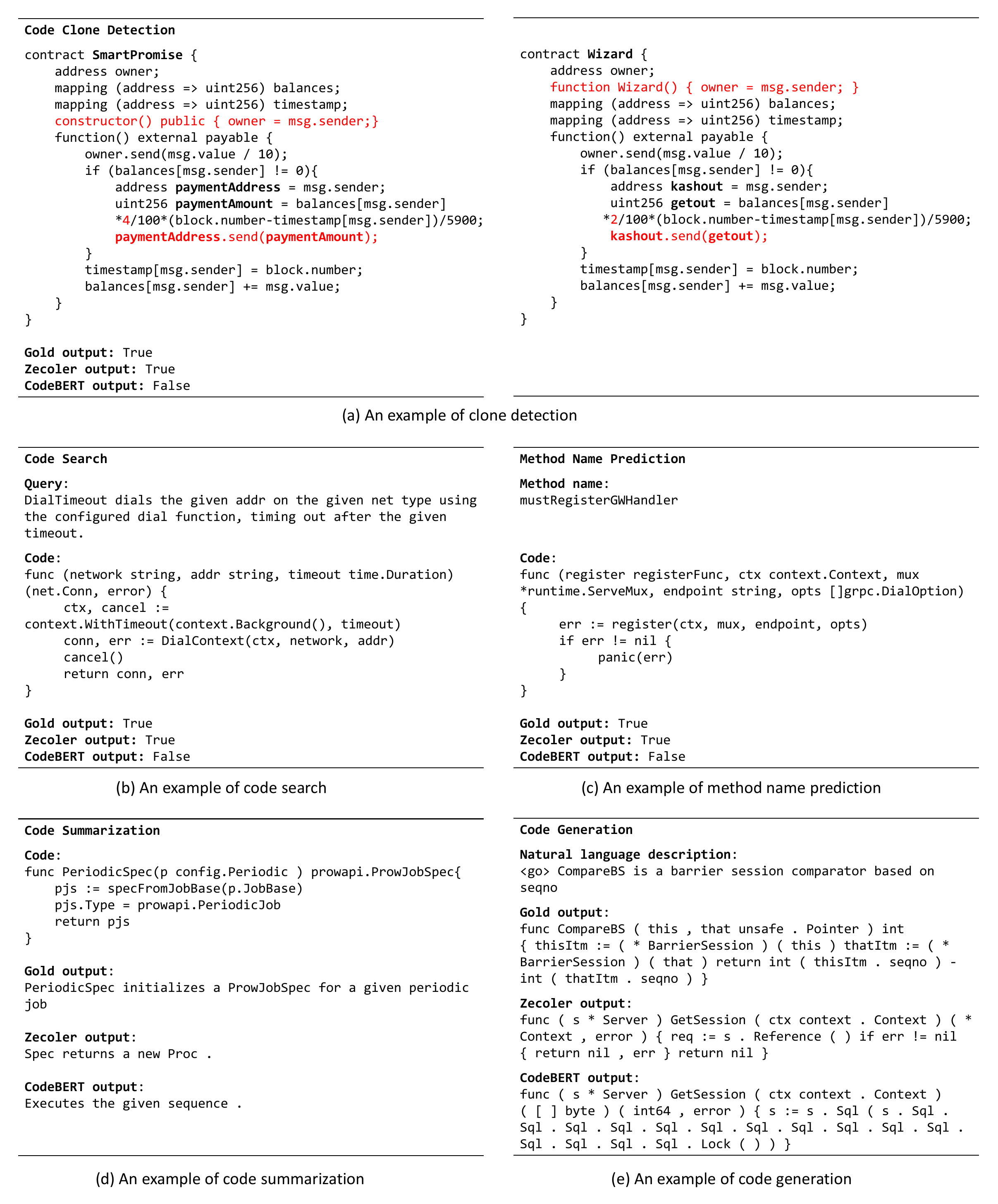}}
    \caption{Examples of five classification and generative tasks. \approach can successfully predict the correct answer while CodeBERT cannot.}
    \label{fig:case_study4}
\end{figure}


To demonstrate the effectiveness of our approach, we qualitatively analyze concrete examples of five code intelligence tasks by \approach and CodeBERT.

Figure~\ref{fig:case_study4}(a) shows the results of clone detection for two code snippets in Solidity. They are similar in functionality while being different in key words and structures. For example, ``paymentAddress'' and ``kashout'' (highlighted in red) are two equivalent keywords in the two snippets. Because the two words are both domain specific, baseline models such as CodeBERT can hardly detect the clone without prior knowledge. Comparatively, \approach successfully detects the clone by reusing prior knowledge from PLMs using prompt learning. 


\new{Figure~\ref{fig:case_study4}(b) shows an example of code search. The query searches for programs that dial with a time-out limit. The target code involves a Go specific API ``context.WithTimeout" which can hardly be recognized by CodeBERT. Thanks to the zero-shot learning of \approach, our approach successfully recognizes both the domain-specific API and the query, hence gives a correct answer. }

\new{Figure~\ref{fig:case_study4}(c)
compares the results by \approach and CodeBERT in the method name prediction task of Go. 
The ``panic" in the code is a language-specific API which interrupts the program when an error occurs. As can be seen, CodeBERT can hardly comprehend the function ``panic" in zero-shot scenario, while \approach can fully understand it and predict the correct method name.}


\new{Figure~\ref{fig:case_study4}(d) presents an example of code summarization. Compared with the summary generated by CodeBERT, the one generated by \approach contains more precise keywords such as ``\rev{Spec}". Also, the meaning of ``new" in its output is similar to that of ``initializes" in the expected one. In contrast, CodeBERT is not capable of grasping the meaning of the code snippet, and thus the generated summary is far from the expected output.}

\new{Figure~\ref{fig:case_study4}(e) compares
the generated code in Go for the query ``CompareBS is a barrier session comparator based on seqno".
The \approach generates code with correct syntax especially the return and if statements, while many repeated error tokens such as ``Sql." can be found in the code generated by CodeBERT. }

These examples demonstrate the superiority of \approach in
zero-shot code representation learning. The prompts in \approach cast the underlying meaning in the PLM to downstream tasks, which help large PLMs capture text and code semantics even without training data.



\section{\rev{Discussions}}
 \subsection{\rev{Limitations}}

 \rev{The effectiveness of zero-shot learning is only verified on Go and Solidity, thus it remains questionable whether Zecoler is also effective for other low-level languages such as RUST and C++.    
 However, we believe our approach can be adapted to more low-level languages including RUST and C++. Because Zecoler gains common knowledge about programming languages through pre-training while requiring no supervision of target languages. That means it has high transparency when adapted to a wider range of other languages.}
    
 \rev{The dataset used for evaluating \approach is relatively smaller (e.g., 5000) compared to the conventional settings when we evaluate fine-tuning methods. Hence, \approach could hardly reach the same accuracy of fine-tuning when the data is sufficiently available. Many works \citep{prefixtuning, softprompt} related to prompt also aware of this and use ways such as limiting data set size or freezing PLM parameters for training to highlight the effectiveness of prompt in the scenario of few-shot learning or lightweight deployment.}

 \rev{Another limitation lies in the time cost for prompt-tuning. Although prompt tuning involves updating much fewer parameters compared with fine-tuning, it remains challenging to search for the global optima in gradient descent. Our results indicate that prompt tuning consumes about 30\% more time cost than fine-tuning. }

\subsection{Threats to Validity}

\textit{Internal Validity.}
Our approach is built upon CodeBERT. Although CodeBERT is the most popular PLM for learning code representations, other PLMs for code such as CodeT5 (with an encoder-decoder architecture) and Codex (unidirectional Transformers) may have different results. However, we argue that our approach is independent on the PLM architecture itself since we merely modify the format of input and output of the PLM. 


\new{Prompts have a small amount of trainable parameters and may lead to unstable results. Although we alleviate this problem by employing an LSTM, we still observe that the prompt representations fall into local minima in gradient descend. We have to train the model repeatedly with different random seeds for the best automatically-generated prompts. We leave the dynamic optimization of prompts for our future work.}

\smallskip\noindent\textit{External Validity.}
In our work, the classification downstream tasks are assumed to be binary classifications. Hence, we represent the binary answers using two candidate words. Hence, more candidate words for multi-class classification tasks remain to be investigated. The candidate words are manually selected and can be searched to find the most suitable ones. 
\new{We can further represent the candidate words as trainable vectors just like the prompts in our approach.} 


\section{Related Work}
\label{sec:related}
\subsection{Learning Code Representations}

As the core prerequisite for many code intelligence tasks, learning code representations has been extensively explored in software engineering~\citep{codexglue}. 
Broadly, typical approaches in learning code representations can be classified into three categories, including unsupervised for general languages, supervised for specific tasks, and few-shot learning.

The most typical category of work lie in the unsupervised approaches such as code2vec~\citep{code2vec}, code2seq~\citep{code2seq}, and InferCode~\citep{infercode}. Code2vec and code2seq aggregate representations of each path in AST (abstract syntax tree) based on attention. InferCode predicts subtrees automatically identified from the contexts of an AST in a self-supervised manner.
These methods directly learn code representation from AST paths. They utilize the word embedding techniques in natural language processing and incorporate them with semantic and syntax information in program source code.
The limitation of these methods is the lack of adaptions to downstream tasks. 
The learned code vectors are fixed and cannot be fine-tuned on downstream tasks.
Furthermore, these methods are purely trained on code, thus are unsuitable for NL-PL tasks such as code search.

To improve the performance of downstream tasks, researchers have also resorted to task-oriented supervised learning methods~\citep{codenet}. 
For example, for code clone detection task, \citet{FangLS0S20} caught the similarity of semantics between two code snippets using a supervised deep learning model, which pays attention to caller-callee relationships and learns the hidden syntactic and semantic features of source codes. 
\citet{ZhangHZWLS21} disentangled the representation of semantic and syntax with AST and GAN (generative adversarial network), then used only semantic representation to detect code clone.
For code search task,
\citet{gu2018deepcs} proposed a code representation model named CODEnn to learn semantic representations of code snippets through jointly embedding with comments,
\citet{HaldarWXH20} designed a multi-perspective cross-lingual neural framework, and \citet{Liw20} learned code-query interactions. 
\citet{ZhangCLP21} proposed a hybrid code representation learning approach to resolve program dependence and semantics for predicting method name.
\citet{YangCZSZ21} learned a unified vector representation of both methods and bug reports for method-level fault localization.
\citet{ZhouLSD019} constructed a graph neural network to learn semantic representation of code to identify the vulnerable functions.
\citet{WangL21a} proposed AST Graph Attention Block to capture different dependencies in the AST graph for representation learning in code completion.
These models are trained for the specific downstream tasks, which achieve good performance but lack generality to support multiple tasks with one single model. 

The aforementioned methods require a large scale corpus to train the code representation model.
To alleviate this problem, pre-trained programming language models are proposed such as CodeBERT~\citep{codebert} and CodeT5~\citep{codet5}. It is a fine-tuning based few-shot program learning paradigm: PLMs learn a vast amount of knowledge from large scale unlabelled corpora in the pre-training phase, and achieve state-of-the-art accuracy in the fine-tuning phase with a small amount of labelled task-specific data.
This gives PLMs the basic generalization ability to handle a wide range of downstream tasks well. Task adaption through fine-tuning adds extra knowledge of specific tasks to PLMs and improves the performance.
However, in this paradigm, the gap between the pre-training phase and the downstream task can be significant: the objectives are different, and for the downstream tasks, we usually need to introduce new parameters.


To the best of our knowledge, our \approach is the first zero-shot learning method for code representation. 
\approach follows a prompt-based learning paradigm for task adaption.
Prompt learning makes it possible for downstream tasks to take the same format as the pre-training objectives and require no new parameters. 
By narrowing the gap between the two phases, deploying the PLMs on specific tasks becomes much easier with little training data.


\subsection{Prompt-based Learning}
\label{sec:zeroshotlearning}


As a promising method for zero-shot learning, a growing number of prompt-based learning approaches~\citep{abs210713586} have been proposed in recent years. For example, \citet{pet} proposed PET which transforms the classification task into an MLM task and uses prompt to elicit knowledge from PLM. But the prompt is manually crafted and hard to select the most suitable words for it. \citet{autoprompt} proposed AutoPrompt which automatically searches prompt words discretely using gradient signals in the target task. Although discrete searching retains the semantic of prompt, it also cannot find out the most precise prompts for machine models. For solving this problem, \new{instead of using prompt for solving downstream tasks directly, \citet{abs-2210-14803} proposed a method that uses prompt for filtering training dataset and trains the model more efficiently. But the quality of dataset is hard to control, which makes the accuracy of downstream tasks low. \citet{ArcoVK22} tackled this challenge with combinations of multiple prompts for more robust prompt template. But this way relies on human selection and still can not make sure to find the best prompt.}
\citet{prefixtuning} proposed Prefix-Tuning which optimizes a continuous task-specific vector prepended to every layer of the Transformer in PLM and freezes the PLM for saving computation cost. Prefix-Tuning demonstrates superb performance but it only focuses on natural language generative tasks. 
\new{\citet{nomorefinetune} empirically evaluated the usage and effect of prompt tuning in code intelligence tasks including defect prediction, code summarization, and code translation. The results show that prompt tuning outperforms fine-tuning in full data, and also shows great potential in low-resource scenarios.}

Comparatively, \approach optimizes the prompt vectors in continuous space instead of discrete words or human-writing, making the prompt more suitable for PLMs to understand and more efficient for extracting knowledge. Moreover, \approach is the first prompt method \new{for zero-shot code representation that can generalize
to various code understanding and generative tasks}.


\section{Conclusion}
\label{sec:conclusion}
In this paper, we propose \approach, a novel approach for  zero-shot code representation learning via prompt tuning. \approach improves traditional pre-trained programming language models by introducing prompt into code representation learning. 
Experiments show that \approach outperforms baseline models in \new{both code understanding and generative tasks} under zero-shot settings. 
Code representations learned by \approach also demonstrate good generalizability for low-resource programming languages.

In the future, \new{we will investigate our approach in more software engineering tasks, with other pretrained models such as CodeT5 and Codex.
We will also consider more characteristics of source code such as syntactic structures in the design of prompt and verbalizer. }

\section*{Data Availability}
Our source code and experimental data are publicly available at: https://github. com/ChrisCN97/zecoler/tree/emse.

\section*{Acknowledgments}
This research is supported by National Natural Science Foundation of China (Grant No. 62232003, 62102244), CCF-Tencent Open Research Fund (RAGR20220129).

\section*{Conflict of Interest}
The authors declared that they have no conflict of interest.

\bibliographystyle{spbasic}
\bibliography{references.bib}


\end{document}